\documentclass[amsmath,amssymb,aps,prb,twocolumn,superscriptaddress, reprint]{revtex4-2} 
\usepackage{svg}
\usepackage{graphicx}
\usepackage{dcolumn}
\usepackage{bm}
\usepackage{comment}
\usepackage{multirow}
\usepackage{array}
\usepackage{braket}
\usepackage{tabularx}
\usepackage{amsmath}
\usepackage{xcolor}
\usepackage[normalem]{ulem}
\usepackage{hyperref}
\usepackage{soul}
\setulcolor{red}
\usepackage{booktabs}
\usepackage{siunitx}
\DeclareUnicodeCharacter{2212}{\ensuremath{-}}

\begin{document}

\title{Sr$_2$NbO$_4$: A $4d$ analogue of the layered perovskite Sr$_2$VO$_4$}

\author{Leonid S. Taran}
\email{leonidtaran97@gmail.com}
\affiliation{M. N. Mikheev Institute of Metal Physics, Ural Branch of Russian Academy of Sciences, 620137 Yekaterinburg, Russia}

\author{Anastasia E. Lebedeva}
\affiliation{Institute of Physics and Technology, Ural Federal University, 620002 Yekaterinburg, Russia}

\author{Sergey V. Streltsov}
\affiliation{M. N. Mikheev Institute of Metal Physics, Ural Branch of Russian Academy of Sciences, 620137 Yekaterinburg, Russia}
\affiliation{Institute of Physics and Technology, Ural Federal University, 620002 Yekaterinburg, Russia}

\date{\today}

\begin{abstract}
This work focuses on the layered perovskite Sr$_2$NbO$_4$, a 4\textit{d} analogue of Sr$_2$VO$_4$, which remains an unsolved puzzle with a possible intriguing hidden magnetic order. Using density functional theory (DFT) calculations, we demonstrate the robust thermodynamic stability and exfoliability of Sr$_2$NbO$_4$, suggesting potential applications as a 2D material. Imperfect Fermi surface nesting indicates instabilities that may drive symmetry lowering, charge/orbital density waves, or superconductivity.  Dynamical mean-field theory (DMFT) calculations reveal moderate mass renormalization $(m^*/m\sim1.3)$ and an itinerant character of magnetism with strong longitudinal spin fluctuations. The exchange interaction is dominated by in-plane ferromagnetic coupling with much weaker interlayer antiferromagnetic exchange. 
\end{abstract}

\maketitle

\section{INTRODUCTION}

Layered perovskites have become one of the most studied crystal structures since the discovery of high-temperature superconductivity in cuprates. The mysterious superconductivity in another layered perovskite, Sr$_2$RuO$_4$, has been a subject of continuous study for more than 30 years~\cite{maeno1994,rice1995,mackenzie2003,barber2019}, while its cousin, Ca$_2$RuO$_4$, has become a playground for investigations of the orbital-selective Mott transition~\cite{anisimov2002,de2009,de2011}. The absence of long-range magnetic order in Sr$_2$VO$_4$~\cite{Cyrot1990,sugiyama2014} has kindled interest in possible hidden orders, such as, for example, the ordering of magnetic octupoles or other higher-order multipoles~\cite{Jackeli2009a,Kim2017b,Igoshev2024}. The flame of this interest later spread to $d^1$ and $d^2$ double perovskites, see e.g.~\cite{chen2010,Maharaj2020,Pourovskii2021,Takayama2021}, but the puzzle of Sr$_2$VO$_4$ remains unsolved.

Interestingly, layered nickelates with a single electron in the transition metal (TM) $e_g$ shell were initially considered analogues of superconducting cuprates, which have one hole in the $e_g$ shell~\cite{anisimov1999}. However, only recently progress in synthesis has led to the discovery of the long-sought superconductivity in layered nickelates~\cite{li2019,hepting2020}. Similarly, Sr$_2$IrO$_4$, with a single hole in the TM $t_{2g}$ shell, is considered a potential gateway to cuprate-like superconductivity~\cite{mitchell2015}. All of this makes the study of layered perovskites with one $t_{2g}$ electron particularly compelling. Indeed, it has been suggested that uniaxial pressure could drive Sr$_2$VO$_4$ into a superconducting regime~\cite{arita2007}.

In the present paper, Sr$_2$NbO$_4$, the $4d$ analogue of Sr$_2$VO$_4$ with a single electron in the $t_{2g}$ shell, is considered. This material has been scarcely studied experimentally. It was first synthesized in 1974 and characterized as ``the crystal type of perovskites like Sr$_2$TiO$_4$''\cite{Kasimov1974}, which corresponds to the $I4/mmm$ space group\cite{Villars2023:sm_isp_sd_1502124}. The lattice parameter was reported with a precision of a single digit after the decimal point\cite{Kasimov1974}. Magnetic measurements are controversial: Isawa and Nagano reported that the magnetic susceptibility ($\chi$) follows the Curie-Weiss law but did not present experimental data~\cite{Isawa2001}. In contrast, Ref.~\cite{nakamura1994} indicates that the low-temperature part of $\chi(T)$ is close to zero and shows no temperature dependence.

The electronic properties of Sr$_2$NbO$_4$ are also unclear. The temperature dependence of resistivity, $\rho(T)$, suggests semiconducting behavior over a wide temperature range, but the experiments were conducted on a powder sample, where grain boundary effect may be significant~\cite{Isawa2001}. In addition,  detailed studies of the sister-material Sr$_2$VO$_{4+\delta}$ showed that $\rho(T)$ strongly depends on oxygen content and variation of $\delta$ from 0.15 to -0.15 changes character from insulating to metallic~\cite{ueno2014}. Additionally, thermoelectric power $S(T)$ measurements reveal several anomalies (not observed in resistivity)~\cite{Isawa2001}, with the absolute value of $S(T)$ at low temperatures being consistent with a metallic phase.

We studied chemical stability of Sr$_2$NbO$_4$ by calculating enthalpy of formation, present details of obtained crystal structure characterized by tetragonal symmetry in accordance with available experimental data and show that this material is potentially exfoliable. Analysis of the electronic and magnetic structure suggests that it is closer to itinerant regime with moderate renormalization of the band structure close to the Fermi level. In turn, the imperfect nesting observed in non-magnetic DFT calculations indicates instability, which can lead to a further lowering of symmetry (e.g., down to orthorhombic), the formation of a charge or orbital density wave, or superconductivity.

\section{Methods}
All calculations were performed using the Perdew-Burke-Ernzerhof version of the generalized gradient approximation (GGA) \cite{Perdew1997} employing the \textsc{vasp} code \cite{Kresse1996}. The cutoff energy for the plane-wave basis set was chosen to be 440 eV. The stopping criterion for electronic self-consistency was set to 10$^{-7}$ eV. Integration over the Brillouin zone was carried out using a $6 \times 6 \times 2$ Monkhorst-Pack $k$-point mesh \cite{Monkhorst1976}. The Wigner-Seitz radii for niobium, strontium, and oxygen were chosen to be 1.270, 2.138, and 0.820 \AA, respectively.
\begin{figure}[b!]
\centering
\includegraphics[width=0.5\columnwidth]{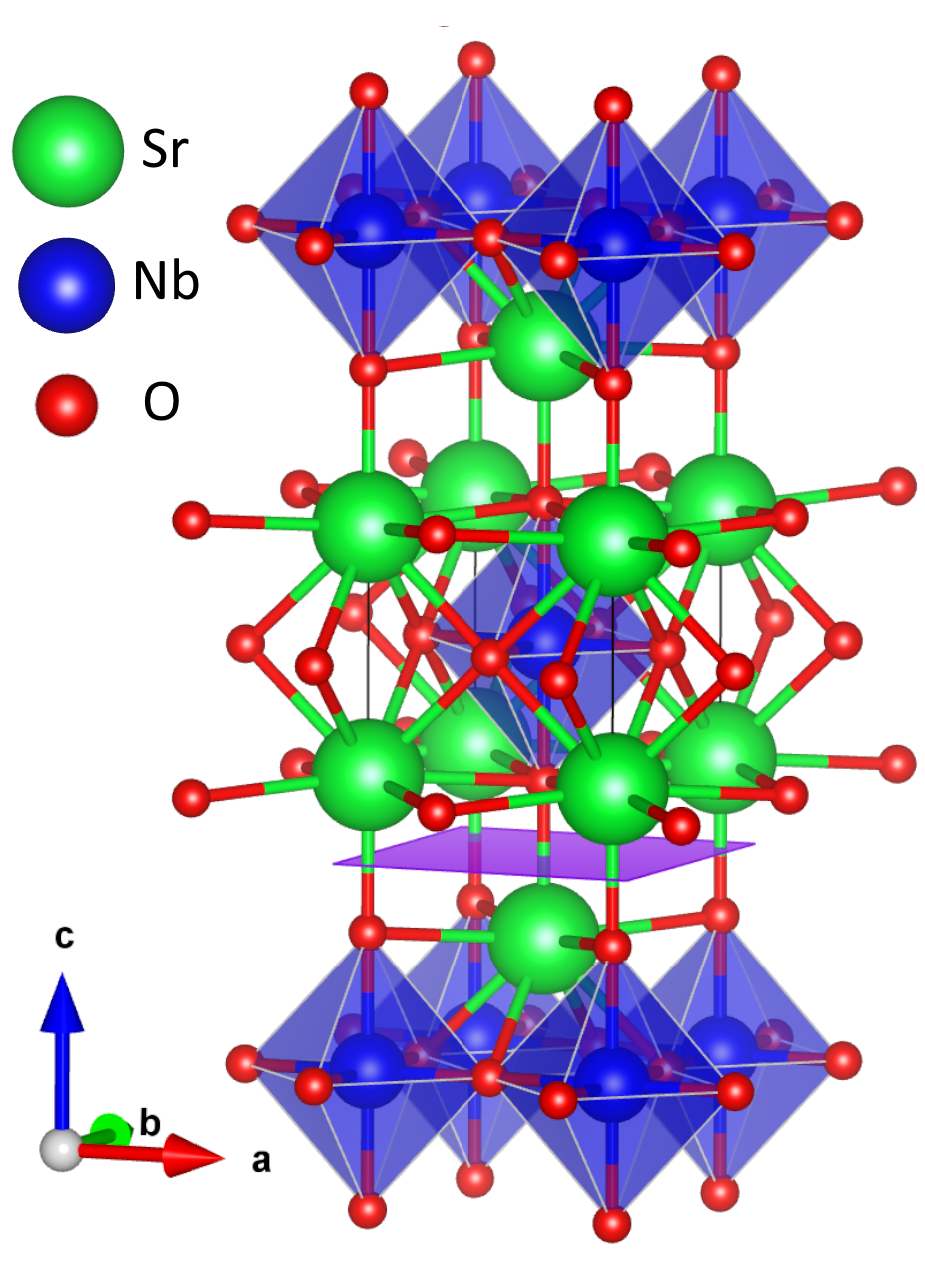}
    \caption{Polyhedral representation of the Sr$_2$NbO$_4$ conventional bulk cell with the position of the (001) cleavage plane (violet). Blue octahedra consist of oxygen ligands (red) and niobium atoms in the center (blue). Polyhedra with strontium atoms (green) are not considered.}
    \label{Fig: Structure}
\end{figure}

The considered crystal structure was subjected to a full relaxation procedure, which included optimization of atomic positions, cell shape, and volume, using the conjugate gradient algorithm \cite{Press1986}. The force tolerance criterion for the convergence of atomic positions was set to 10$^{-3}$ eV/\AA, while convergence criterion for the total energy was chosen to be 10$^{-6}$ eV. The final atomic structure was visualized with {\sc vesta} \cite{VESTA}.

The on-site Coulomb interaction was included using the rotationally invariant DFT+U approach introduced by Liechtenstein {\it et al.} \cite{Liechtenstein1995}. The Hubbard $U$ parameter for niobium was calculated to be $U_\mathrm{Nb} = 2$ eV using the linear response approach proposed by Cococcioni and de Gironcoli \cite{Cococcioni2005}. The effective intra-atomic exchange interaction was set to $J^\mathrm{Nb}_\mathrm{H} = 0.5$ eV. \cite{Sasioglu2011}.

The paramagnetic dynamical-mean field theory calculations were performed using AMULET code\cite{AMULET} with the segment version of the continuous-time quantum Monte-Carlo (QMC) solver~\cite{gull2011}. Small 3$\times$3 Hamiltonian corresponding to $t_{2g}$ orbitals of Nb was obtained by calculating wannier functions \cite{Marzari1997,Marzari2012} using Wannier90 code ~\cite{Mostofi2014}. While in Sec.~\ref{DFT} we discuss hopping parameters for a few nearest neighbors, the non-interacting Hamiltonian for DMFT calculations includes all possible hoppings. Coulomb interaction term was considered to be in the generalized Kanamori (GK) form.~\cite{Georges2013} Note, that this is an approximation, since real symmetry of the impurity problem is lower than cubic (tetragonal). The spin-orbit coupling was not included to DFT+DMFT calculations as it introduces off-diagonal terms to non-interacting Hamiltonian. More than $10^6$ QMC sweeps were used at each DMFT iteration to guarantee ergodicity. 300 slices on imaginary time were used in calculations and imaginary-time analytical continuation was performed by Pad\'e approximation.


\section{Crystal structure}
\begin{figure}[t!]
\centering
\includegraphics[width=1\columnwidth]{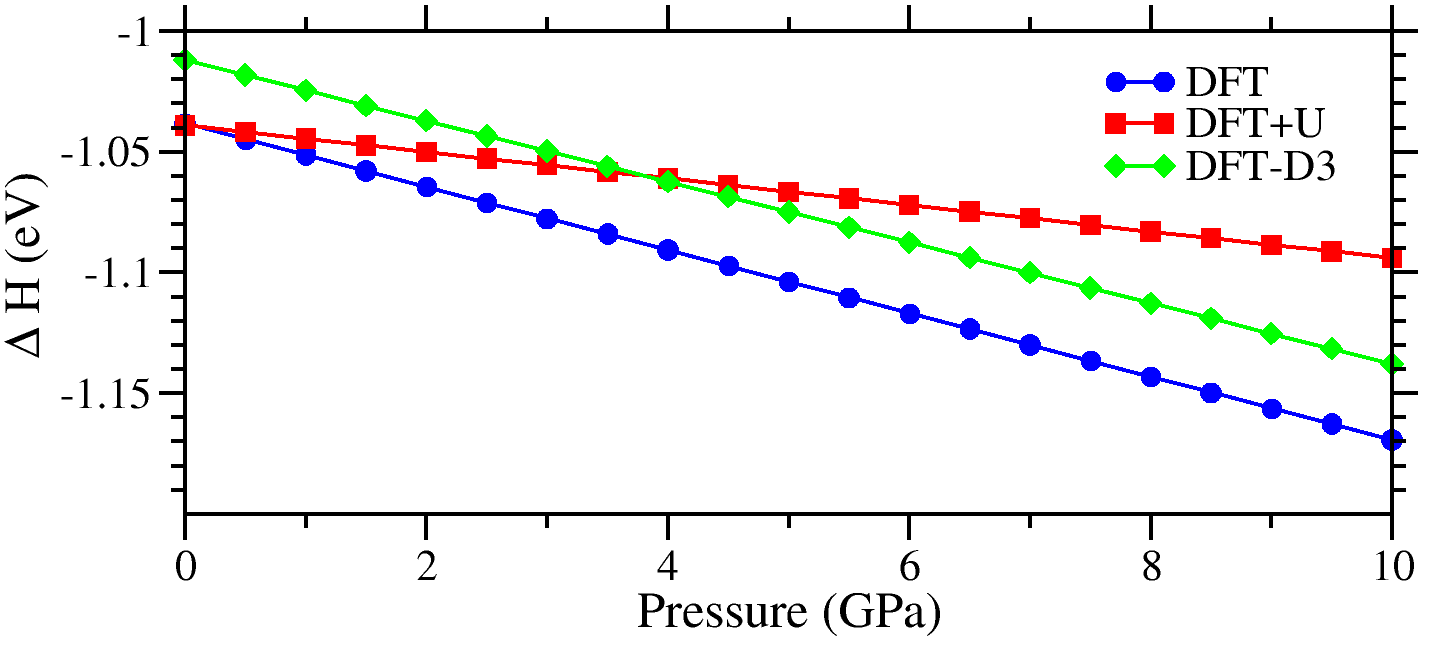}
    \caption{Relative enthalpy of Sr$_2$NbO$_4$ as obtained within DFT (blue), DFT+U (red), and DFT-D3 taking into account van-der-Waals correction (green) approaches.}
    \label{Fig: Relative enthalpy}
\end{figure}

The initial structural model was based on the Sr$_2$VO$_4$ structure reported by Cyrot \textit{et al.} \cite{Cyrot1990}. In this model, V atoms were replaced by Nb, followed by full structural relaxation. The obtained lattice parameters and interatomic distances are summarized in the Supplemental Material, Table S1, whereas atomic positions are given in Table S2 \cite{Taran2025}. The resulting structure remained tetragonal characterized by $I4/mmm$ space group, which is consistent with conclusions of \cite{Kasimov1974}. The lattice parameters of the relaxed structure were determined to be $a = 4.03869$ \AA~and $c = 12.75256$ \AA. This is in a reasonable agreement with the first publication on Sr$_2$NbO$_4$ synthesis, where $a=4.1$\AA ~(with precision of a single digit after dot) was reported and the structure was characterized to have the same structural type as Sr$_2$TiO$_4$ (space group was not specified)\cite{Kasimov1974}.

The conventional unit cell contains two formula units. The structure corresponds to the Ruddlesden-Popper phase \cite{Ruddlesden1957,Ruddlesden1958}, which has the general formula $A_{n+1}B_nX_{3n+1}$. The NbO$_6$ octahedra are stretched along the \textit{c}-axis: four short Nb-O bonds are 2.02 \AA, while two long bonds are 2.08 \AA. These octahedra share their corners to form a square lattice in the \textit{ab} plane, but do not stack directly on top of each other in the \textit{c} direction\cite{Wells1975}, as illustrated in Fig.~\ref{Fig: Structure}.

Next, to study the thermodynamic stability of Sr$_2$NbO$_4$ we calculated the energy-volume dependence for the primitive cells of the reactants and the final product involved in the chemical reaction, as described by Kasimov \textit{et al.} \cite{Kasimov1974}:  
\begin{eqnarray}
\mathrm{NbO_2} + 2\mathrm{SrO} \xrightarrow{} \mathrm{Sr_2NbO_4}.
\end{eqnarray}

Since the reaction under consideration occurs in vacuum at a temperature range of 1000--1200$^{\circ}$C, we used the crystal structures of the reagents that are stable at these conditions. For each compound, volume of the primitive cell was varied within $10$\% with respect to its equilibrium value. The energy-volume curves for Sr$_2$NbO$_4$, NbO$_2$, and SrO are shown in Fig. S1 \cite{Taran2025}. The parameters of the equation of state (EOS), including the equilibrium volume $V_0$ and the bulk modulus $B_0$, were determined by fitting the calculated energy-volume data to the third-order Birch–Murnaghan EOS \cite{Birch1947}. The resulting parameters are listed in Table~\ref{table:EOS}. The thermodynamic stability was investigated by comparing the enthalpies of Sr$_2$NbO$_4$ and its components NbO$_2$ and SrO, as shown in Fig. \ref{Fig: Relative enthalpy}. The results indicate that the enthalpy of formation, $\Delta H$, is negative in DFT calculations over a wide range of pressures, including at 0 GPa (ambient pressure), confirming that Sr$_2$NbO$_4$ is thermodynamically stable. Interestingly, both account of the van der Waals interaction or strong Hubbard repulsion via DFT-D3 and DFT+U, respectively, do not change the situation considerably.
\begin{table}[b!]
    \centering
    \caption{Parameters for the equation of state for Sr$_2$NbO$_4$, NbO$_2$, and SrO. $V_0$ stands for the equilibrium volume, $B_0$ is a bulk modulus.}
    \begin{ruledtabular}
    \begin{tabular}{lccc}
       & Sr$_2$NbO$_4$ & NbO$_2$ & SrO \\ \midrule
        $V_0$, \AA${^3}$ & 208.07 & 71.64 & 140.71 \\ 
        $B_0$, GPa   & 129.23 & 235.61 & 85.33 \\ 
    \end{tabular}
    \end{ruledtabular}
    \label{table:EOS}
\end{table}

Finally, we have calculated the integrated crystal orbital Hamiltonian projection (ICOHP) to get further insight of the chemical bonding (the full list of -ICOHPs is shown in Table S3 of SM \cite{Taran2025}). As expected, the strength of Nb-O bonds ranges from 4.04 to 4.46 eV, which is much large than Sr-O bonding (0.23-0.71 eV). 

A detailed inspection of the crystal structure reveals that the (001) surface has the lowest density of Sr--O bonds. Cleavage along this surface involves breaking two bonds per $ab$ plane of the conventional unit cell, as shown in Fig.~\ref{Fig: Structure}. Using Sr--O bond energies, we estimate the energy required for a cleavage perpendicular to (001) direction  as 1.44 J/m$^2$ (90 meV/\AA$^2$), which places Sr$_2$NbO$_4$ to category of potentially exfoliable materials \cite{Mounet2018}.

\section{A short account on theoretical results on a sister material S\MakeLowercase{r}$_2$VO$_4$}
\begin{figure}[t!]
\centering
\includegraphics[width=1\columnwidth]{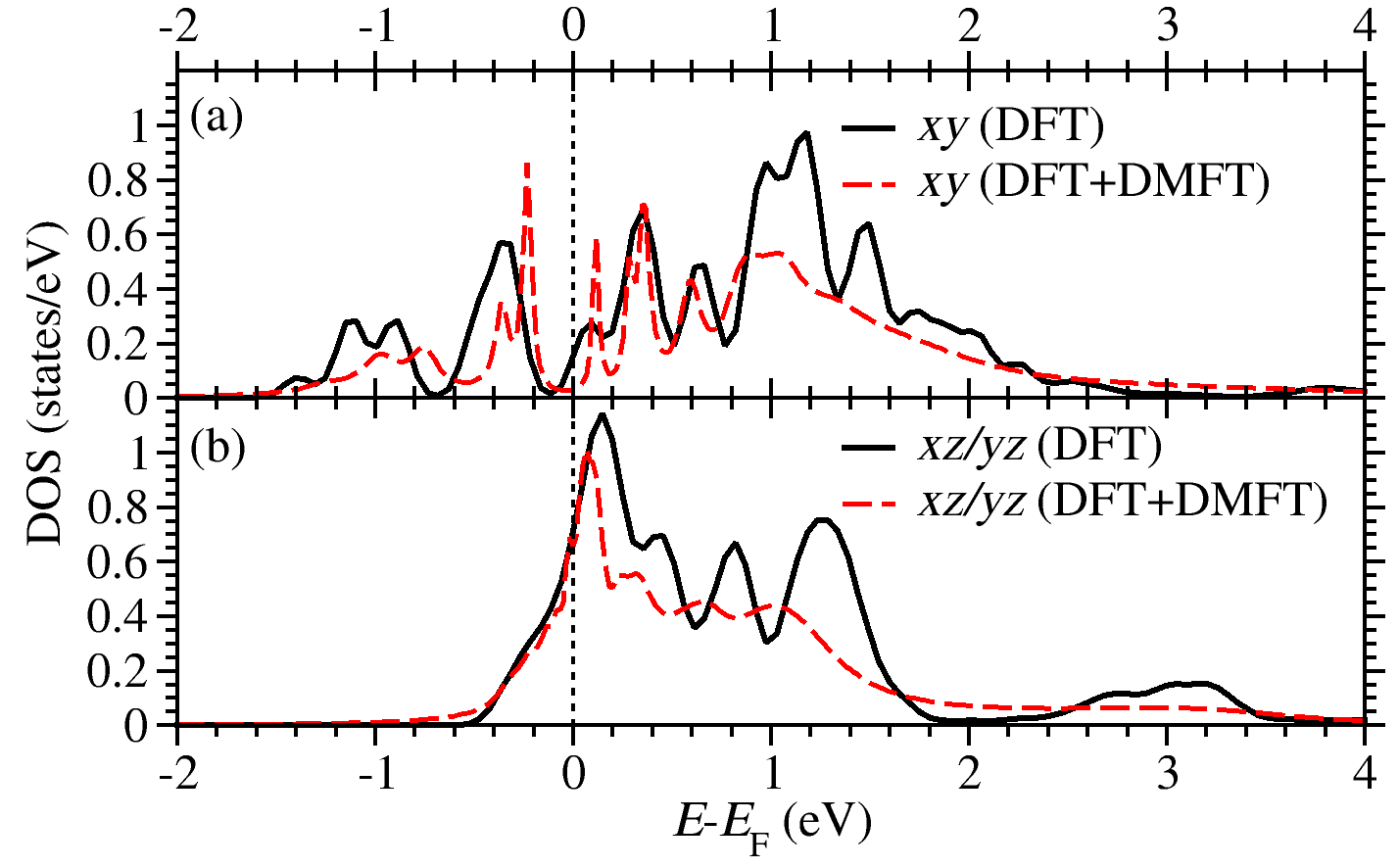}
    \caption{Partial DOS (solid black) obtained by DFT and spectral functions (dash red lines) calculated by DFT+DMFT at $T=100K$ 
    for Nb \textit{xy} (a) and \textit{xz/yz} (b) orbitals in a primitive cell.}
    \label{Fig: DFT(+DMFT) DOS}
\end{figure}

Let us briefly recall the current theoretical understanding of the sister material Sr$_2$VO$_4$ before discussing the results of our \textit{ab initio} calculations for Sr$_2$NbO$_4$. The tetragonal elongation (in $c$ direction) typical for layered perovskites leads to a situation where the $xy$ orbital is expected to lie higher in energy than the other two $t_{2g}$ orbitals. Consequently, one might anticipate localization of a single electron in insulating Sr$_2$VO$_4$ \cite{matsuno2005a} on these two orbitals. According to the Jahn-Teller effect, this should result in a lowering of the crystal structure symmetry and possibly induce orbital order.  However, such distortions have not been experimentally observed so far. Instead, several exotic scenarios have been proposed, incorporating spin-orbit coupling and involving stabilization of the electron on appropriate spin-orbitals to avoid further symmetry lowering \cite{Jackeli2009a,Eremin2011}.

Recent Hartree-Fock calculations of a three-band model with a single electron on a square lattice, taking into account strong Coulomb repulsion and spin-orbit coupling shed  some light on Sr$_2$VO$_4$~\cite{Igoshev2024}. In particular, these calculations demonstrated that the previously proposed state characterized by alternating $j^z_{\text{eff}} = \pm 3/2$ orbitals~\cite{Eremin2011} never stabilizes, as it exhibits a significantly higher total energy (here, $j^z_{\text{eff}}$ denotes the $z$-projection of the effective total moment for the $t_{2g}$ electrons, described by an effective orbital moment $l_{\text{eff}} = 1$ and spin $s = 1/2$). Instead, several competing phases emerge: (1) an antiferromagnetic state with electron localization at the $xy$ orbital (AFM-$xy$), favored when both the tetragonal splitting of the $t_{2g}$ levels and the intra-atomic Hund's coupling $J_H$ are relatively small; (2) a ferromagnetic state with an antiferro-orbital ground state (FM-AFO$_{xy/1}$), stabilized for more realistic values of $J_H$; and (3) an orbitally entangled state (AFO-eO), where conventional (dipolar) magnetic and orbital orders are suppressed~\cite{Jackeli2009a}. The choice of a particular state is dictated not only by interaction parameters but also by details of the non-interacting Hamiltonian, which determines the tetragonal splitting and hopping parameters.

\section{Non-magnetic DFT \label{DFT} for S\MakeLowercase{r}$_2$N\MakeLowercase{b}O$_4$}
\begin{figure}[t!]
\centering
\includegraphics[width=1\columnwidth]{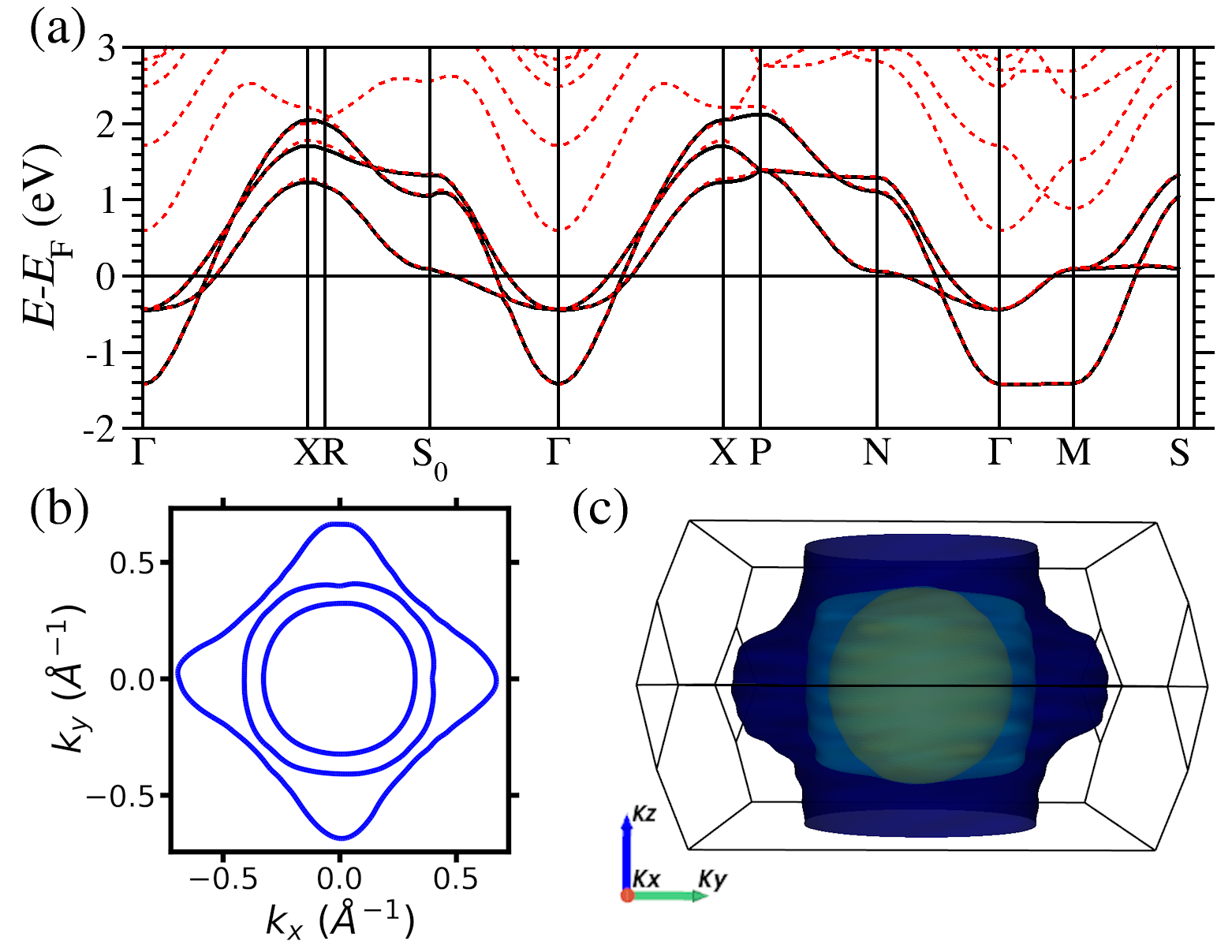}
    \caption{(a) Electronic band structure obtained in the non-magnetic DFT calculations is shown by dashed red lines. Solid black lines is the band structure corresponding to the projected on Wannier functions used for plotting the Fermi surface, DMFT and estimation of hopping integrals. 2D Fermi for $k_{z}=0$ and 3D Fermi surface obtained in non-magnetic DFT calculations are shown in (b) and (c) respectively. Dark blue, light blue and yellow surfaces correspond to  the $xy$, $yz$, and $xz$ bands, respectively. }
    \label{Fig: bands&fermi}
\end{figure}

The electronic band structure of Sr$_2$NbO$_4$ obtained in non-magnetic DFT calculations is presented in Figs.~\ref{Fig: DFT(+DMFT) DOS}, \ref{Fig: bands&fermi}(a) and Fig. S2 of SM \cite{Taran2025}. There are Nb $4d$ states at the Fermi level with the $xy$ band wider than the degenerate $xz/yz$ bands (local coordinate systems with axes pointing to ligands is used here and below). This is because in layered perovskite structure the $xy$ orbital overlaps (via O 2$p$ orbitals) with all 4 its neighbors, while $xz/yz$ only with two of them. Interestingly, there is a pseudogap close to the Fermi level for the $xy$ band and a small hole doping can place it exactly at the Fermi level. A sister material Sr$_2$VO$_4$ does not have such a pseudogap~\cite{arita2007}. Another interesting feature of Sr$_2$NbO$_4$ is a Van Hove singularity at $\sim 100$ meV above the Fermi level (note, that M-S is not along $k_z$), see \ref{Fig: bands&fermi}(a). This resembles situation in Sr$_2$RuO$_4$~\cite{barber2019}.

In order to compare electronic structures of Sr$_2$VO$_4$ and Sr$_2$NbO$_4$ more accurately and place Sr$_2$NbO$_4$ on the phase diagram obtained by the many-body Hartree-Fock calculations including interplay between spin and orbital degrees of freedom we calculated effective hopping parameters for Sr$_2$NbO$_4$. 
The Wannier function formalism was applied for this \cite{Marzari1997,Marzari2012}. Maximally localized Wannier functions were constructed for the Nb $t_{2g}$ orbitals. The final spreads were 3.3 \AA$^2$ ($xy$) and 4.1 \AA$^2$ ($xz/yz$).

Comparison of the projected and initial band structures are presented in Fig.~\ref{Fig: bands&fermi}(a). The hopping integral between the $xz,yz$ orbitals was found to be $t_{xz,xz}=t_{yz,yz}= 357$ meV, which is 1.4 times larger than in Sr$_2$VO$_4$. This increase is due to a larger principal quantum number. Interestingly, nearest neighbor hopping here is very close to the one in the effective single-band model in cuprates~\cite{hybertsen1990}. The hopping integrals for the next nearest neighbors were found to be $t'_{xz,xz}=t'_{yz,yz}$  = 9 meV  and  $t'_{xz,yz} =t'_{yz,xz}$ = 7 meV. We emphasize, that the interlayer 3rd and in-plane 4th neighbors are not negligible: $t''_{xz,xz} =t''_{yz,yz}$= 34 meV, $t''_{xz,yz}=t''_{yz,xz}$= 38 meV, and $t'''_{xz,xz} =t'''_{yz,yz}$= 35 meV. As can be seen, $t''$ become order of magnitude smaller than those for nearest neighbors, but not vanishingly small. The complete set of obtained hopping integrals, along with their graphical representation, is presented in the Supplementary Material (Table S4 and Figure S3).

The crystal-field splitting within the $t_{2g}$ shell was found to be $\Delta_{CF}=90$ meV  with the $xy$ orbital lying higher than the $xz/yz$ doublet ($\Delta_{CF}/t=0.25$). The spin-orbit coupling constant for Nb$^{4+}$ is $\sim$93 meV~\cite{Abragam} ($\lambda/t=0.26$) and Hubbard repulsion is of order of 2 eV ($U/t \sim 6$). Taking into account that Hund's exchange is expected to be $J_H = 0.5$ eV ($J_H/t=1.4$) \cite{Sasioglu2011},  Sr$_2$NbO$_4$ can appear to be in the region of two ferromagnetic states - antiferro-orbital orbital entangled (AFO-eO) or antiferro-orbital with alternating $xz/yz$ orbitals in the phase diagram presented in Ref.~\cite{Igoshev2024}. It has to be mentioned at this point that application of the Hartree-Fock approximation used in Ref.~\cite{Igoshev2024} is rather questionable in case of Sr$_2$NbO$_4$ (our results presented in the next section show that it turns out to be metallic for the crystal structure under consideration), but surprisingly the ferromagnetic ground state is consistent with DFT+U results.

The three $t_{2g}$ bands form three distinct Fermi surface sheets, as shown in Fig.~\ref{Fig: bands&fermi}(b,c). While the $xz$ and $yz$ bands produce nearly spherical surfaces, the $xy$ orbital generates a cylindrical sheet in k-space, which expands into a deformed rectangle near $k_z=0$.
This implies an imperfect nesting with $\vec Q = (\pi, \pi, 0)$, which can potentially result in an instability leading to, e.g., magnetism, the formation of the charge density wave or superconductivity.

\begin{figure}[t!]
\centering
\includegraphics[width=1\columnwidth]{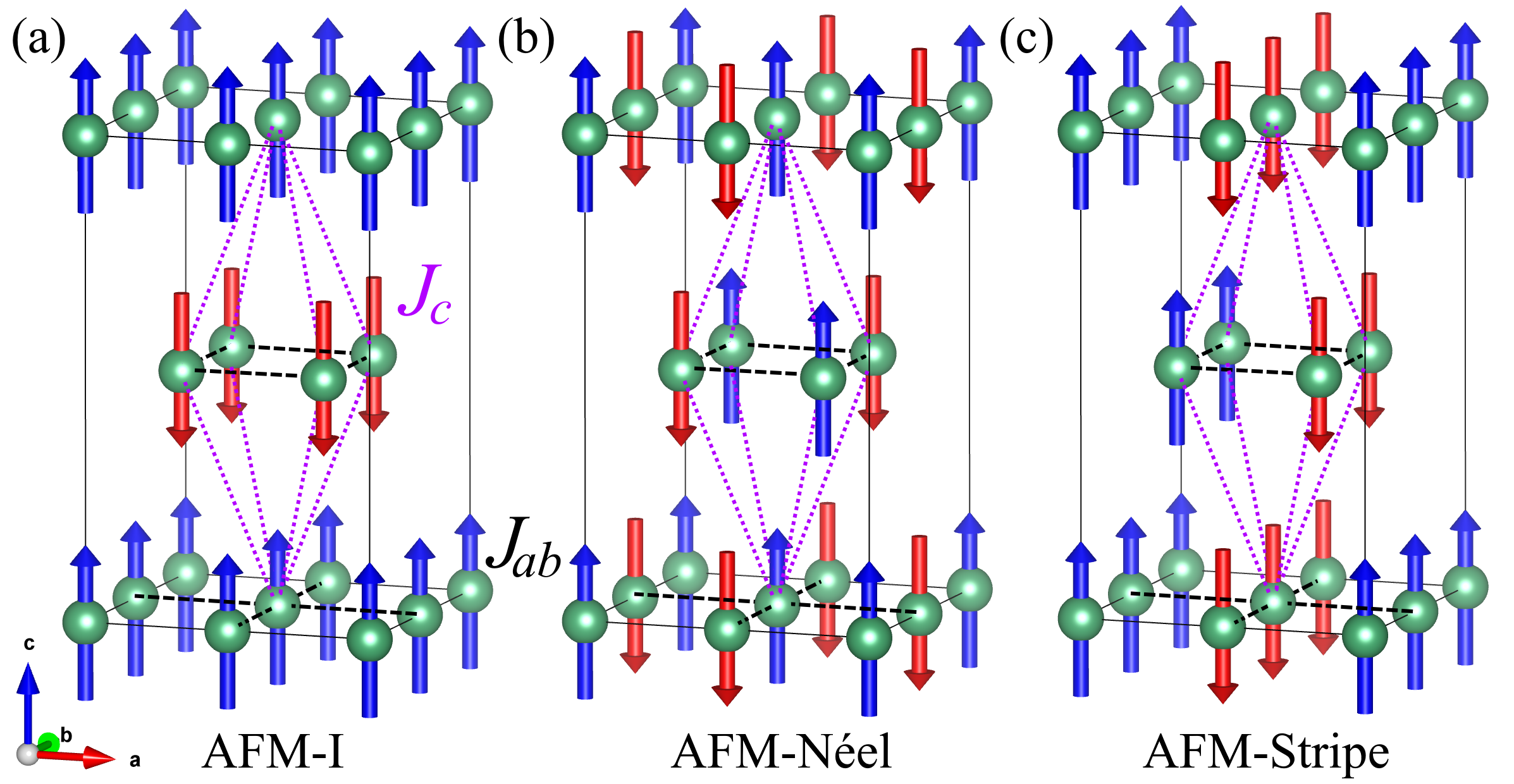}
    \caption{(a) $2 \times 2 \times 1$ supercell used in magnetic calculations; (a-c) three antiferromagnetic configurations of Sr$_2$NbO$_4$: layered AFM or AFM-I, AFM-Néel and AFM-Stripe, respectively. The nonmagnetic Sr and O ions are ignored to give a clear picture of the magnetic configurations in the Nb sublattice. In-plane $J_{ab}$ and interlayer $J_c$ exchange couplings denoted by black and purple dash lines, respectively.}
    \label{Fig: Magnetic conf.}
\end{figure}

\section{ Account of correlation effects}

\subsection{\label{GGAU} DFT+U results}
We start with the simplest approach treating correlation effects on the static mean-field level by DFT+U approach and calculate 4 spin configurations to find the magnetic ground state. These are ferromagnetic (FM), and three antiferromagnetic states (AFM-I, AFM-N\'eel, and AFM-Stripe), see Fig.  \ref{Fig: Magnetic conf.}. The optimized (in non-magnetic GGA) crystal structure was used in these calculations.

\begin{table}[b!]
    \centering
    \caption{The calculated total energies of Sr$_2$NbO$_4$ magnetic configurations within the DFT+U per formula unit. All the energies are given relative to that of the ground state in meV.}
    \begin{ruledtabular}
    \begin{tabular}{lcc}
        Configuration & $J_{H}=0.35$ eV & $J_{H}=0.5$ eV  \\ \midrule
        NM & 82.5 & 83.9 \\ 
        FM & 8.8 & 8.8 \\ 
        AFM-I & 0 & 0 \\ 
        AFM-N\'eel & 81.0 & 80.2 \\ 
        AFM-Stripe & 15.2 & 18.0 \\
    \end{tabular}
    \end{ruledtabular}
    \label{table:GS}
\end{table}

Resulting total energies are presented in Table \ref{table:GS}. One can see that AFM-I with antiferromagnetic interlayer order turns out to be the ground state. The next is FM order signaling, that the interlayer exchange coupling is AFM (and small). The Heisenberg model defined as 
\begin{eqnarray}
H = \sum_{i \ne j} J \vec S_i \vec S_j,
\end{eqnarray}
was used to describe the magnetic interactions in Sr$_2$NbO$_4$. For simplicity restricting ourselves by the nearest neighbor exchanges we obtain AFM $J_c=26$ K and FM $J_{ab}=-440$ K (it is worth noting that the exchange coupling between further neighbors can be important; see corresponding calculations for sister compound Sr$_2$VO$_4$~\cite{Kim2017b}). The FM order obtained in DFT+U calculations is consistent with the model phase diagram presented in Ref.~\cite{Igoshev2024}. Magnetic moments on Nb were found to be $\sim$0.5$\mu_B$, strongly reduced from ionic values due to hybridization effects (typical for $4d/5d$ transition metal compounds, see, e.g., \cite{streltsov2013}). 

\begin{figure}[t!]
\centering
\includegraphics[width=1\columnwidth]{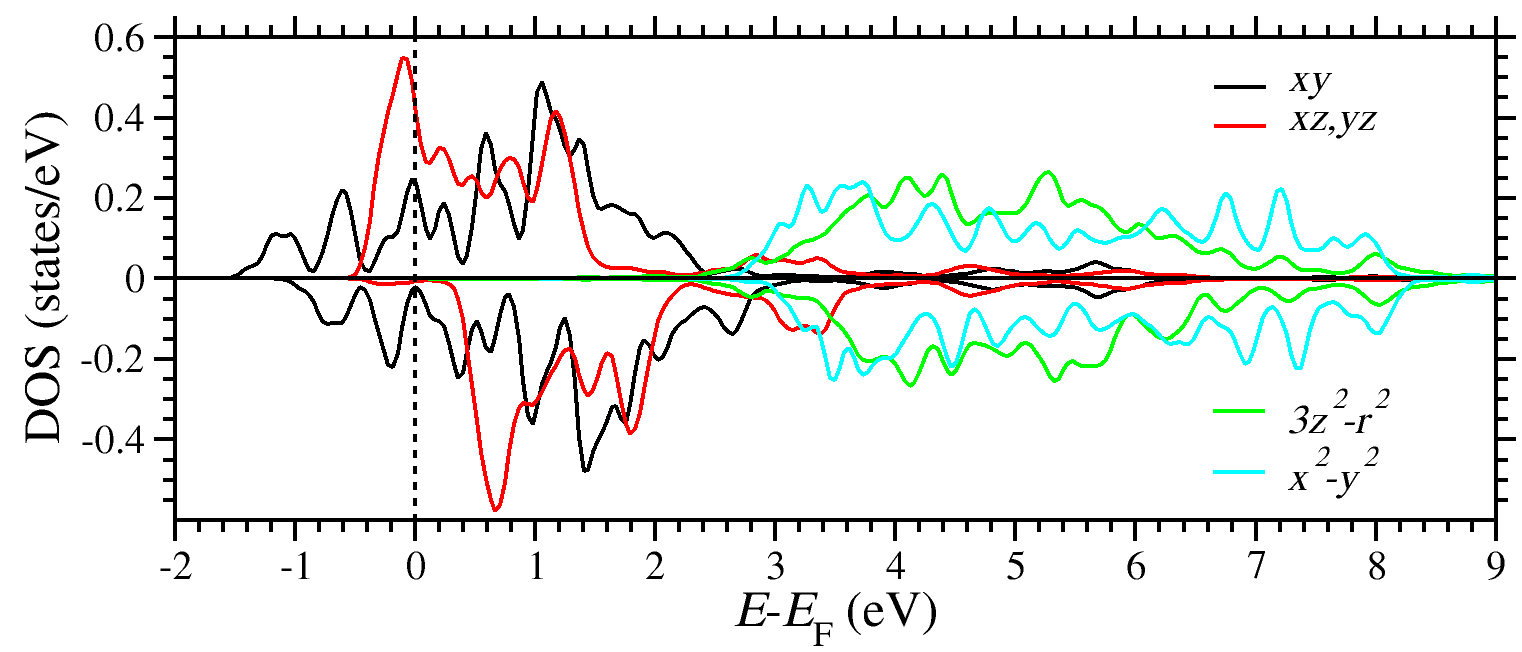}
    \caption{Decomposed $d$-orbital Nb partial density of states (DOS) with spin up obtained  in DFT+U calculation. Colored lines represent DOS projected on the $xy$ (black), $xz$ and $yz$ (red), $3z^2-r^2$ (green),  and $x^2-y^2$ (cyan). Energy is given with respect to E$_F$ (vertical dash line).}
    \label{Fig: DFT+U DOS}
\end{figure}

The total and projected density of states (DOS), presented in Fig.~\ref{Fig: DFT+U DOS} and Fig. S4 of the SM \cite{Taran2025} for the ground state AFM-I order, show that the valence O-$p$ states lie in the energy range from -8 to -3 eV. The Nb-$t_{2g}$ states extend from -1 to 2 eV, crossing the Fermi level, indicate the metallic character of Sr$_2$NbO$_4$ even at the static mean-field level. Most probably, this is related to the large width of the $t_{2g}$ band, exceeding 2 eV, which is comparable to the Hubbard $U$. To consider correlation effects more accurately in this situation, we applied the DFT+DMFT method, the results of which are discussed in the next sub-section. Interestingly account of the spin-orbit coupling does not change DOS of Sr$_2$NbO$_4$ (or magnetic moments), see Fig. S4 \cite{Taran2025}. This is most probably related to metallic ground state with delocalized electrons.

Taking into account the nesting-driven instability suggested by non-magnetic DFT calculations, which may lead to an enlargement of the unit cell, we performed additional structural optimization within the DFT+U+SOC approach. However, this did not result in the stabilization of the antiferro-orbital $xz/yz$ ordering.
More detailed studies considering the lowering of the crystal structure to orthorhombic symmetry may be required to explore this possibility.

\subsection{\label{DMFT} DFT+DMFT results}

\begin{figure}[t!]
\centering
\includegraphics[width=1\columnwidth]{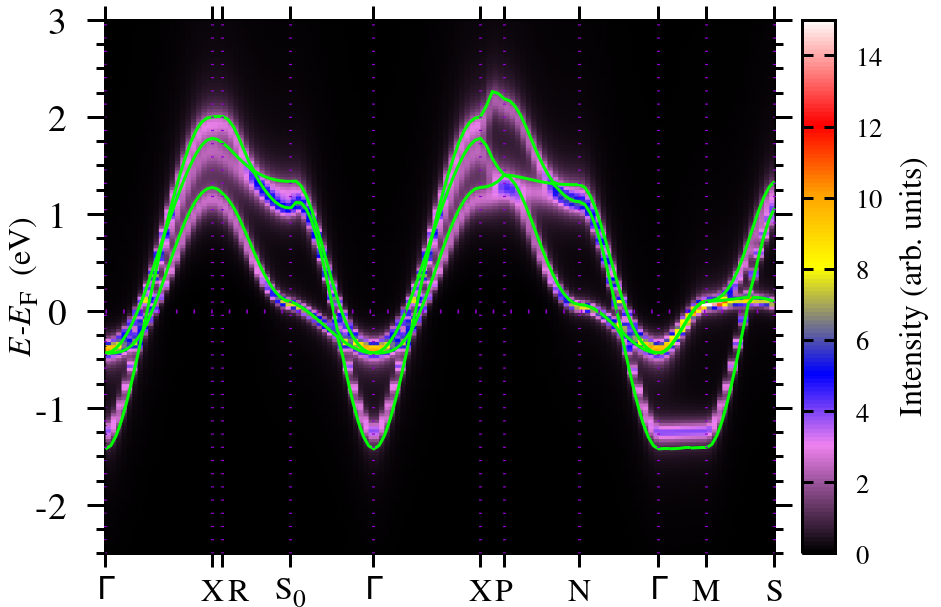}
    \caption{Electronic spectrum as obtained in paramagnetic DFT+DMFT (color map) and non-magnetic DFT (green) calculations. Energy is indicated relative to E$_F$.}
    \label{Fig: DFT(+DMFT) spectrum}
\end{figure}

The DFT+DMFT calculations were performed for $T=100$K in the paramagnetic state, since in spite of sizable exchange coupling within the $ab$ plane, $J_{ab}$, a layered structure implies substantial suppression temperature of the magnetic ordering. One can see from Fig.~\ref{Fig: DFT(+DMFT) DOS} that this more accurate (than DFT+U) approach shows only a subtle renormalization of the DFT electronic structure. The most important one is the shift of the Fermi level to the pseudogap for the $xy$ orbital, which, however, does not transform to the real gap. In Fig.~\ref{Fig: DFT(+DMFT) spectrum} band structures obtained in DFT and DFT+DMFT approaches are compared. The mass renormalization is $m^*/m$ is 1.32 for the $xz/yz$ and 1.18 for $xy$ bands evidencing weak correlation effects.

\begin{figure}[t!]
\centering
\includegraphics[width=1\columnwidth]{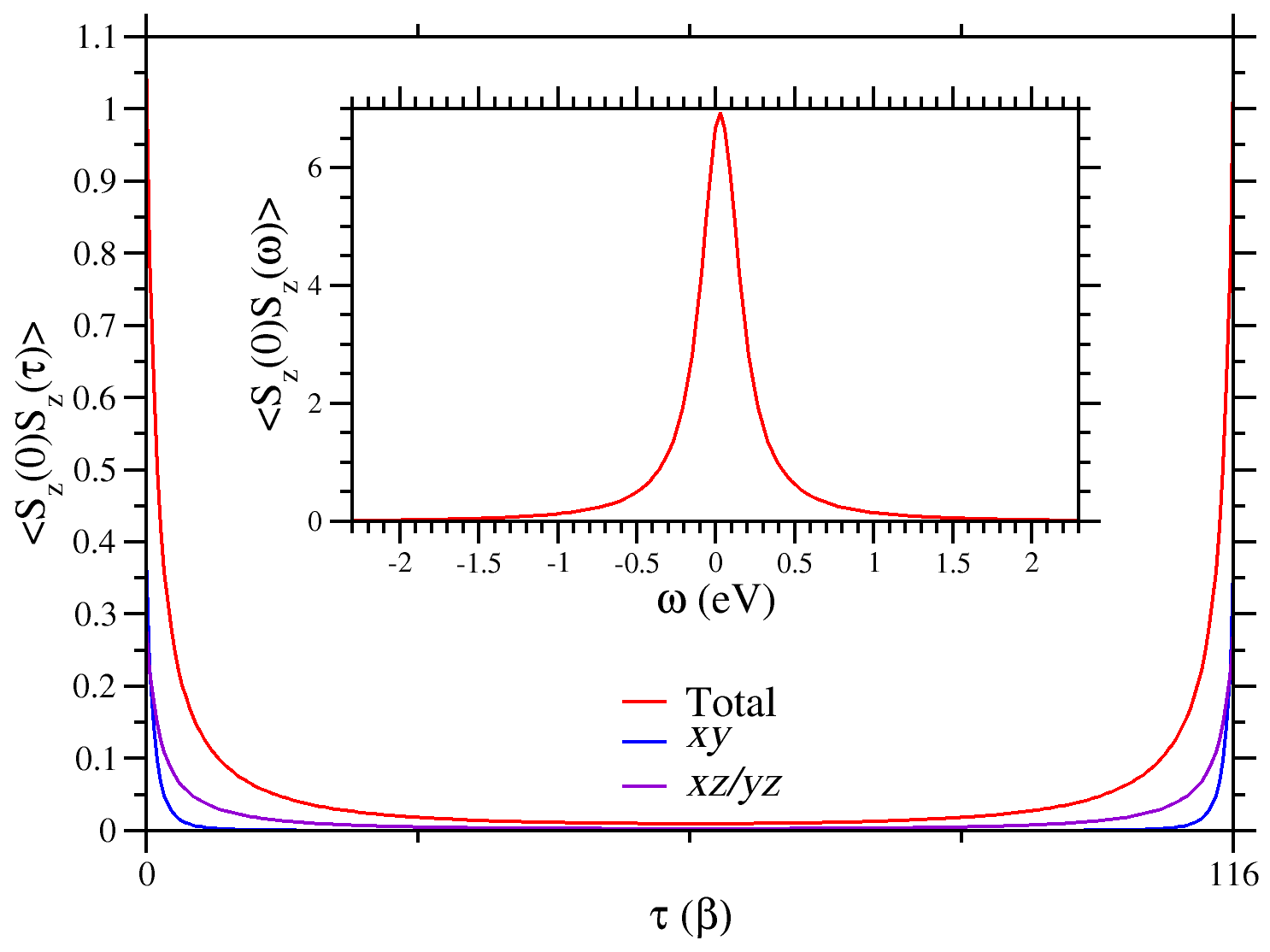}
    \caption{Imaginary time, $\tau$, and frequency, $\omega$, (inset) dependence of the local spin correlator in Sr$_2$NbO$_4$ calculated by DFT+DMFT for $T=100 K$ ($\beta \sim 116$ eV$^{-1}$). Instant magnetic moment was found to be  $g\sqrt {\langle S^2_z \rangle} =1.08 \mu_B$.}
    \label{Fig: spin correlator}
\end{figure}

This agrees with a strong imaginary time dependence of the on-site spin-spin correlator $\langle \hat S^z(0) \hat S^z(\tau) \rangle$, see Fig. \ref{Fig: spin correlator}. Half-width ($\Delta \varepsilon$) of its Fourier transform  to real frequencies presented in inset of Fig.~\ref{Fig: spin correlator} allows to calculate lifetime ($\Delta \tau \Delta \varepsilon \ge h $) of spin, which in our case turns out to be 35 fs. This is somewhat longer than what was calculated for a textbook example of itinerant magnet ZrZn$_2$ (25 fs~\cite{katanin2023}), but is still shorter than lifetimes in typical materials with localized electrons such as, e.g., $\alpha$-Fe ($\varepsilon \sim 0.09$ eV, lifetime is $\sim 50$ fs)\cite{igoshev2013} or LiZn$_2$Mo$_3$O$_8$ with lifetime of  $60$ fs (for total spin per Mo$_3$ triangle)\cite{streltsov2025}. 

It has to be mentioned that this result does not contradict to Ref.~\cite{Isawa2001}, where observation of the Curie-Weiss law was reported. Indeed, there are many itinerant magnets still following the Curie-Weiss law \cite{moriya2012}. More puzzling is nearly zero magnetic susceptibility reported in \cite{nakamura1994} at Helium temperatures (if only this is not a signature of a superconducting phase resulting to diamagnetism).

Finally,  it has to be noted that we employed Hubbard $U=2$ eV, as determined by our linear response calculations for the full $4d$ shell, while the DMFT calculations were performed only for the $t_{2g}$ subshell. The Hubbard $U$ for the more localized Nb $t_{2g}$ orbitals could be slightly larger, although values of $U \sim 1-3$ eV are typically reported for Nb in the literature~\cite{Vaugier2012,moore2024}. A somewhat larger value of $U=6$ eV yielded a quasiparticle residue of $Z \sim 0.6$ and consequently a larger effective mass of $m^* \sim 1.67$ for Sr$_2$NbO$_4$ in Ref.~\cite{paul2019}.

\section{Discussions and Conclusions}
To sum up, we theoretically studied Sr$_2$NbO$_4$, a $4d$ analogue of Sr$_2$VO$_4$, which is famous for the unsolved puzzle of its hidden magnetic order. Experimental literature on Sr$_2$NbO$_4$ provides only limited information on its crystal structure and contradictory data on its magnetic properties. We show that the enthalpy of formation for this materials is large and negative independent on the calculation method used, which guaranties that this layered Nb oxide exists. Moreover, this material is potentially exfoliable with a cleavage energy of 1.44 J/m$^2$ and can be used for construction of a 2D square lattice.

Interestingly, the Fermi surface shows imperfect nesting with $\vec Q = (\pi, \pi, 0)$, suggesting an instability that could result in a lowering of the tetragonal symmetry of the crystal lattice, for example, due to the formation of a charge (or orbital) density wave or superconductivity. According to the phase diagram of the three-band Hubbard model~\cite{Igoshev2024}, the parameters characterizing the electronic structure of Sr$_2$NbO$_4$ suggest the possible formation of an antiferro-orbital order with alternating $xz/yz$ orbitals or the orbital-entangled ground state proposed for Sr$_2$VO$_4$ associated with a hidden magnetic order~\cite{Jackeli2009a}. Stabilization of the antiferro-orbital order is consistent with imperfect nesting observed in non-magnetic DFT calculations. It can result in narrowing of the $t_{2g}$ band and would additionally work for the insulating conductivity reported in one of two experimental studies (on polycrystalline samples).  

However, direct optimization of the crystal structure in the present DFT+U(+SOC) calculations does not lead to such antiferro-orbital ordering. Nor does it support the stabilization of the (N\'eel) antiferromagnetic ground state associated with the wave vector $\vec{Q} = (\pi, \pi, 0)$. Instead, it reveals a strong ferromagnetic exchange interaction with nearest neighbors in the $ab$ plane and an antiferromagnetic interaction between layers.

Account of dynamical correlation effects via DFT+DMFT approach shows that mass renormalization is rather small, $m^*/m \sim 1.3$, which does not strongly affect the electronic structure close to the Fermi level. From point of view of magnetism this material is closer to itinerant magnets with strong longitudinal fluctuations and not very large lifetime of spin excitations (comparable to ZrZn$_2$).

The present results motivate further experimental studies of Sr$_2$NbO$_4$, including a detailed investigation of its magnetic properties, a search for possible lowering of crystal symmetry or superconductivity on the square lattice of Nb layers. Such studies can be promising, considering the recent progress in the synthesis and investigation of layered superconducting nickelates with a very similar lattice.

\begin{acknowledgments}
Authors are grateful to P. Igoshev, S. Skornyakov, and K. Kugel for stimulating discussions and to A. Poteryaev and A. Shorikov for their assistance with the technical implementation of DMFT calculations.

We thank the Ministry of Science and Higher Education of the Russian Federation for support through funding the Institute of Metal Physics.
\end{acknowledgments}

\appendix

\bibliography{main}

\begin{thebibliography}{58}%
\makeatletter
\providecommand \@ifxundefined [1]{%
 \@ifx{#1\undefined}
}%
\providecommand \@ifnum [1]{%
 \ifnum #1\expandafter \@firstoftwo
 \else \expandafter \@secondoftwo
 \fi
}%
\providecommand \@ifx [1]{%
 \ifx #1\expandafter \@firstoftwo
 \else \expandafter \@secondoftwo
 \fi
}%
\providecommand \natexlab [1]{#1}%
\providecommand \enquote  [1]{``#1''}%
\providecommand \bibnamefont  [1]{#1}%
\providecommand \bibfnamefont [1]{#1}%
\providecommand \citenamefont [1]{#1}%
\providecommand \href@noop [0]{\@secondoftwo}%
\providecommand \href [0]{\begingroup \@sanitize@url \@href}%
\providecommand \@href[1]{\@@startlink{#1}\@@href}%
\providecommand \@@href[1]{\endgroup#1\@@endlink}%
\providecommand \@sanitize@url [0]{\catcode `\\12\catcode `\$12\catcode `\&12\catcode `\#12\catcode `\^12\catcode `\_12\catcode `\%12\relax}%
\providecommand \@@startlink[1]{}%
\providecommand \@@endlink[0]{}%
\providecommand \url  [0]{\begingroup\@sanitize@url \@url }%
\providecommand \@url [1]{\endgroup\@href {#1}{\urlprefix }}%
\providecommand \urlprefix  [0]{URL }%
\providecommand \Eprint [0]{\href }%
\providecommand \doibase [0]{https://doi.org/}%
\providecommand \selectlanguage [0]{\@gobble}%
\providecommand \bibinfo  [0]{\@secondoftwo}%
\providecommand \bibfield  [0]{\@secondoftwo}%
\providecommand \translation [1]{[#1]}%
\providecommand \BibitemOpen [0]{}%
\providecommand \bibitemStop [0]{}%
\providecommand \bibitemNoStop [0]{.\EOS\space}%
\providecommand \EOS [0]{\spacefactor3000\relax}%
\providecommand \BibitemShut  [1]{\csname bibitem#1\endcsname}%
\let\auto@bib@innerbib\@empty
\bibitem [{\citenamefont {Maeno}\ \emph {et~al.}(1994)\citenamefont {Maeno}, \citenamefont {Hashimoto}, \citenamefont {Yoshida}, \citenamefont {Nishizaki}, \citenamefont {Fujita}, \citenamefont {Bednorz},\ and\ \citenamefont {Lichtenberg}}]{maeno1994}%
  \BibitemOpen
  \bibfield  {author} {\bibinfo {author} {\bibfnamefont {Y.}~\bibnamefont {Maeno}}, \bibinfo {author} {\bibfnamefont {H.}~\bibnamefont {Hashimoto}}, \bibinfo {author} {\bibfnamefont {K.}~\bibnamefont {Yoshida}}, \bibinfo {author} {\bibfnamefont {S.}~\bibnamefont {Nishizaki}}, \bibinfo {author} {\bibfnamefont {T.}~\bibnamefont {Fujita}}, \bibinfo {author} {\bibfnamefont {J.}~\bibnamefont {Bednorz}},\ and\ \bibinfo {author} {\bibfnamefont {F.}~\bibnamefont {Lichtenberg}},\ }\bibfield  {title} {\bibinfo {title} {Superconductivity in a layered perovskite without copper},\ }\href@noop {} {\bibfield  {journal} {\bibinfo  {journal} {Nature}\ }\textbf {\bibinfo {volume} {372}},\ \bibinfo {pages} {532} (\bibinfo {year} {1994})}\BibitemShut {NoStop}%
\bibitem [{\citenamefont {Rice}\ and\ \citenamefont {Sigrist}(1995)}]{rice1995}%
  \BibitemOpen
  \bibfield  {author} {\bibinfo {author} {\bibfnamefont {T.}~\bibnamefont {Rice}}\ and\ \bibinfo {author} {\bibfnamefont {M.}~\bibnamefont {Sigrist}},\ }\bibfield  {title} {\bibinfo {title} {Sr$_2${R}u{O}$_4$: an electronic analogue of 3{H}e?},\ }\href@noop {} {\bibfield  {journal} {\bibinfo  {journal} {Journal of Physics: Condensed Matter}\ }\textbf {\bibinfo {volume} {7}},\ \bibinfo {pages} {L643} (\bibinfo {year} {1995})}\BibitemShut {NoStop}%
\bibitem [{\citenamefont {Mackenzie}\ and\ \citenamefont {Maeno}(2003)}]{mackenzie2003}%
  \BibitemOpen
  \bibfield  {author} {\bibinfo {author} {\bibfnamefont {A.~P.}\ \bibnamefont {Mackenzie}}\ and\ \bibinfo {author} {\bibfnamefont {Y.}~\bibnamefont {Maeno}},\ }\bibfield  {title} {\bibinfo {title} {The superconductivity of {S}r$_2${R}uo$_4$ and the physics of spin-triplet pairing},\ }\href@noop {} {\bibfield  {journal} {\bibinfo  {journal} {Reviews of Modern Physics}\ }\textbf {\bibinfo {volume} {75}},\ \bibinfo {pages} {657} (\bibinfo {year} {2003})}\BibitemShut {NoStop}%
\bibitem [{\citenamefont {Barber}\ \emph {et~al.}(2019)\citenamefont {Barber}, \citenamefont {Lechermann}, \citenamefont {Streltsov}, \citenamefont {Skornyakov}, \citenamefont {Ghosh}, \citenamefont {Ramshaw}, \citenamefont {Kikugawa}, \citenamefont {Sokolov}, \citenamefont {Mackenzie}, \citenamefont {Hicks} \emph {et~al.}}]{barber2019}%
  \BibitemOpen
  \bibfield  {author} {\bibinfo {author} {\bibfnamefont {M.~E.}\ \bibnamefont {Barber}}, \bibinfo {author} {\bibfnamefont {F.}~\bibnamefont {Lechermann}}, \bibinfo {author} {\bibfnamefont {S.~V.}\ \bibnamefont {Streltsov}}, \bibinfo {author} {\bibfnamefont {S.~L.}\ \bibnamefont {Skornyakov}}, \bibinfo {author} {\bibfnamefont {S.}~\bibnamefont {Ghosh}}, \bibinfo {author} {\bibfnamefont {B.}~\bibnamefont {Ramshaw}}, \bibinfo {author} {\bibfnamefont {N.}~\bibnamefont {Kikugawa}}, \bibinfo {author} {\bibfnamefont {D.~A.}\ \bibnamefont {Sokolov}}, \bibinfo {author} {\bibfnamefont {A.~P.}\ \bibnamefont {Mackenzie}}, \bibinfo {author} {\bibfnamefont {C.~W.}\ \bibnamefont {Hicks}}, \emph {et~al.},\ }\bibfield  {title} {\bibinfo {title} {Role of correlations in determining the van hove strain in {S}r$_2${R}u{O}$_4$},\ }\href@noop {} {\bibfield  {journal} {\bibinfo  {journal} {Physical Review B}\ }\textbf {\bibinfo {volume} {100}},\ \bibinfo {pages} {245139} (\bibinfo {year} {2019})}\BibitemShut {NoStop}%
\bibitem [{\citenamefont {Anisimov}\ \emph {et~al.}(2002)\citenamefont {Anisimov}, \citenamefont {Nekrasov}, \citenamefont {Kondakov}, \citenamefont {Rice},\ and\ \citenamefont {Sigrist}}]{anisimov2002}%
  \BibitemOpen
  \bibfield  {author} {\bibinfo {author} {\bibfnamefont {V.}~\bibnamefont {Anisimov}}, \bibinfo {author} {\bibfnamefont {I.}~\bibnamefont {Nekrasov}}, \bibinfo {author} {\bibfnamefont {D.}~\bibnamefont {Kondakov}}, \bibinfo {author} {\bibfnamefont {T.}~\bibnamefont {Rice}},\ and\ \bibinfo {author} {\bibfnamefont {M.}~\bibnamefont {Sigrist}},\ }\bibfield  {title} {\bibinfo {title} {Orbital-selective mott-insulator transition in {C}a$_{2-x}${S}r$_x${R}u{O}$_4$},\ }\href@noop {} {\bibfield  {journal} {\bibinfo  {journal} {The European Physical Journal B-Condensed Matter and Complex Systems}\ }\textbf {\bibinfo {volume} {25}},\ \bibinfo {pages} {191} (\bibinfo {year} {2002})}\BibitemShut {NoStop}%
\bibitem [{\citenamefont {de’Medici}\ \emph {et~al.}(2009)\citenamefont {de’Medici}, \citenamefont {Hassan}, \citenamefont {Capone},\ and\ \citenamefont {Dai}}]{de2009}%
  \BibitemOpen
  \bibfield  {author} {\bibinfo {author} {\bibfnamefont {L.}~\bibnamefont {de’Medici}}, \bibinfo {author} {\bibfnamefont {S.~R.}\ \bibnamefont {Hassan}}, \bibinfo {author} {\bibfnamefont {M.}~\bibnamefont {Capone}},\ and\ \bibinfo {author} {\bibfnamefont {X.}~\bibnamefont {Dai}},\ }\bibfield  {title} {\bibinfo {title} {Orbital-selective mott transition out of band degeneracy lifting},\ }\href@noop {} {\bibfield  {journal} {\bibinfo  {journal} {Physical review letters}\ }\textbf {\bibinfo {volume} {102}},\ \bibinfo {pages} {126401} (\bibinfo {year} {2009})}\BibitemShut {NoStop}%
\bibitem [{\citenamefont {de’Medici}(2011)}]{de2011}%
  \BibitemOpen
  \bibfield  {author} {\bibinfo {author} {\bibfnamefont {L.}~\bibnamefont {de’Medici}},\ }\bibfield  {title} {\bibinfo {title} {Hund’s coupling and its key role in tuning multiorbital correlations},\ }\href@noop {} {\bibfield  {journal} {\bibinfo  {journal} {Physical Review B—Condensed Matter and Materials Physics}\ }\textbf {\bibinfo {volume} {83}},\ \bibinfo {pages} {205112} (\bibinfo {year} {2011})}\BibitemShut {NoStop}%
\bibitem [{\citenamefont {Cyrot}\ \emph {et~al.}(1990)\citenamefont {Cyrot}, \citenamefont {Lambert-Andron}, \citenamefont {Soubeyroux}, \citenamefont {Rey}, \citenamefont {Dehauht}, \citenamefont {Cyrot-Lackmann}, \citenamefont {Fourcaudot}, \citenamefont {Beille},\ and\ \citenamefont {Tholence}}]{Cyrot1990}%
  \BibitemOpen
  \bibfield  {author} {\bibinfo {author} {\bibfnamefont {M.}~\bibnamefont {Cyrot}}, \bibinfo {author} {\bibfnamefont {B.}~\bibnamefont {Lambert-Andron}}, \bibinfo {author} {\bibfnamefont {J.}~\bibnamefont {Soubeyroux}}, \bibinfo {author} {\bibfnamefont {M.}~\bibnamefont {Rey}}, \bibinfo {author} {\bibfnamefont {P.}~\bibnamefont {Dehauht}}, \bibinfo {author} {\bibfnamefont {F.}~\bibnamefont {Cyrot-Lackmann}}, \bibinfo {author} {\bibfnamefont {G.}~\bibnamefont {Fourcaudot}}, \bibinfo {author} {\bibfnamefont {J.}~\bibnamefont {Beille}},\ and\ \bibinfo {author} {\bibfnamefont {J.}~\bibnamefont {Tholence}},\ }\bibfield  {title} {\bibinfo {title} {Properties of a new perovskite oxyde {S}r$_2${VO}$_4$},\ }\href {https://doi.org/https://doi.org/10.1016/S0022-4596(05)80090-8} {\bibfield  {journal} {\bibinfo  {journal} {Journal of Solid State Chemistry}\ }\textbf {\bibinfo {volume} {85}},\ \bibinfo {pages} {321} (\bibinfo {year} {1990})}\BibitemShut {NoStop}%
\bibitem [{\citenamefont {Sugiyama}\ \emph {et~al.}(2014)\citenamefont {Sugiyama}, \citenamefont {Nozaki}, \citenamefont {Umegaki}, \citenamefont {Higemoto}, \citenamefont {Ansaldo}, \citenamefont {Brewer}, \citenamefont {Sakurai}, \citenamefont {Kao}, \citenamefont {Yang},\ and\ \citenamefont {Martin}}]{sugiyama2014}%
  \BibitemOpen
  \bibfield  {author} {\bibinfo {author} {\bibfnamefont {J.}~\bibnamefont {Sugiyama}}, \bibinfo {author} {\bibfnamefont {H.}~\bibnamefont {Nozaki}}, \bibinfo {author} {\bibfnamefont {I.}~\bibnamefont {Umegaki}}, \bibinfo {author} {\bibfnamefont {W.}~\bibnamefont {Higemoto}}, \bibinfo {author} {\bibfnamefont {E.~J.}\ \bibnamefont {Ansaldo}}, \bibinfo {author} {\bibfnamefont {J.~H.}\ \bibnamefont {Brewer}}, \bibinfo {author} {\bibfnamefont {H.}~\bibnamefont {Sakurai}}, \bibinfo {author} {\bibfnamefont {T.-h.}\ \bibnamefont {Kao}}, \bibinfo {author} {\bibfnamefont {H.-d.}\ \bibnamefont {Yang}},\ and\ \bibinfo {author} {\bibfnamefont {M.}~\bibnamefont {Martin}},\ }\bibfield  {title} {\bibinfo {title} {Hidden magnetic order in {S}r$_2${VO}$_4$ clarified with {$\mu$}+sr},\ }\href {https://doi.org/10.1103/PhysRevB.89.020402} {\bibfield  {journal} {\bibinfo  {journal} {Phys. Rev. B}\ }\textbf {\bibinfo {volume} {89}},\ \bibinfo {pages} {020402} (\bibinfo {year} {2014})}\BibitemShut {NoStop}%
\bibitem [{\citenamefont {Jackeli}\ and\ \citenamefont {Khaliullin}(2009)}]{Jackeli2009a}%
  \BibitemOpen
  \bibfield  {author} {\bibinfo {author} {\bibfnamefont {G.}~\bibnamefont {Jackeli}}\ and\ \bibinfo {author} {\bibfnamefont {G.}~\bibnamefont {Khaliullin}},\ }\bibfield  {title} {\bibinfo {title} {Magnetically hidden order of kramers doublets in {D}1 systems: {S}r$_2${VO}$_4$},\ }\href {https://doi.org/10.1103/PhysRevLett.103.067205} {\bibfield  {journal} {\bibinfo  {journal} {Phys. Rev. Lett.}\ }\textbf {\bibinfo {volume} {103}},\ \bibinfo {pages} {067205} (\bibinfo {year} {2009})}\BibitemShut {NoStop}%
\bibitem [{\citenamefont {Kim}\ \emph {et~al.}(2017)\citenamefont {Kim}, \citenamefont {Khmelevskyi}, \citenamefont {Mohn},\ and\ \citenamefont {Franchini}}]{Kim2017b}%
  \BibitemOpen
  \bibfield  {author} {\bibinfo {author} {\bibfnamefont {B.}~\bibnamefont {Kim}}, \bibinfo {author} {\bibfnamefont {S.}~\bibnamefont {Khmelevskyi}}, \bibinfo {author} {\bibfnamefont {P.}~\bibnamefont {Mohn}},\ and\ \bibinfo {author} {\bibfnamefont {C.}~\bibnamefont {Franchini}},\ }\bibfield  {title} {\bibinfo {title} {Competing magnetic interactions in a spin-1/2 square lattice: Hidden order in {S}r$_2${VO}$_4$},\ }\href {https://doi.org/10.1103/PhysRevB.96.180405} {\bibfield  {journal} {\bibinfo  {journal} {Physical Review B}\ }\textbf {\bibinfo {volume} {96}},\ \bibinfo {pages} {1} (\bibinfo {year} {2017})}\BibitemShut {NoStop}%
\bibitem [{\citenamefont {Igoshev}\ \emph {et~al.}(2024)\citenamefont {Igoshev}, \citenamefont {Chizhov}, \citenamefont {Irkhin},\ and\ \citenamefont {Streltsov}}]{Igoshev2024}%
  \BibitemOpen
  \bibfield  {author} {\bibinfo {author} {\bibfnamefont {P.~A.}\ \bibnamefont {Igoshev}}, \bibinfo {author} {\bibfnamefont {D.~E.}\ \bibnamefont {Chizhov}}, \bibinfo {author} {\bibfnamefont {V.~Y.}\ \bibnamefont {Irkhin}},\ and\ \bibinfo {author} {\bibfnamefont {S.~V.}\ \bibnamefont {Streltsov}},\ }\bibfield  {title} {\bibinfo {title} {Spin-orbit coupling induced orbital entanglement in a three-band hubbard model},\ }\href {https://doi.org/10.1103/PhysRevB.110.115110} {\bibfield  {journal} {\bibinfo  {journal} {Physical Review B}\ }\textbf {\bibinfo {volume} {110}},\ \bibinfo {pages} {115110} (\bibinfo {year} {2024})}\BibitemShut {NoStop}%
\bibitem [{\citenamefont {Chen}\ \emph {et~al.}(2010)\citenamefont {Chen}, \citenamefont {Pereira},\ and\ \citenamefont {Balents}}]{chen2010}%
  \BibitemOpen
  \bibfield  {author} {\bibinfo {author} {\bibfnamefont {G.}~\bibnamefont {Chen}}, \bibinfo {author} {\bibfnamefont {R.}~\bibnamefont {Pereira}},\ and\ \bibinfo {author} {\bibfnamefont {L.}~\bibnamefont {Balents}},\ }\bibfield  {title} {\bibinfo {title} {Exotic phases induced by strong spin-orbit coupling in ordered double perovskites},\ }\href@noop {} {\bibfield  {journal} {\bibinfo  {journal} {Physical Review B—Condensed Matter and Materials Physics}\ }\textbf {\bibinfo {volume} {82}},\ \bibinfo {pages} {174440} (\bibinfo {year} {2010})}\BibitemShut {NoStop}%
\bibitem [{\citenamefont {Maharaj}\ \emph {et~al.}(2020)\citenamefont {Maharaj}, \citenamefont {Sala}, \citenamefont {Stone}, \citenamefont {Kermarrec}, \citenamefont {Ritter}, \citenamefont {Fauth}, \citenamefont {Marjerrison}, \citenamefont {Greedan}, \citenamefont {Paramekanti},\ and\ \citenamefont {Gaulin}}]{Maharaj2020}%
  \BibitemOpen
  \bibfield  {author} {\bibinfo {author} {\bibfnamefont {D.~D.}\ \bibnamefont {Maharaj}}, \bibinfo {author} {\bibfnamefont {G.}~\bibnamefont {Sala}}, \bibinfo {author} {\bibfnamefont {M.~B.}\ \bibnamefont {Stone}}, \bibinfo {author} {\bibfnamefont {E.}~\bibnamefont {Kermarrec}}, \bibinfo {author} {\bibfnamefont {C.}~\bibnamefont {Ritter}}, \bibinfo {author} {\bibfnamefont {F.}~\bibnamefont {Fauth}}, \bibinfo {author} {\bibfnamefont {C.~A.}\ \bibnamefont {Marjerrison}}, \bibinfo {author} {\bibfnamefont {J.~E.}\ \bibnamefont {Greedan}}, \bibinfo {author} {\bibfnamefont {A.}~\bibnamefont {Paramekanti}},\ and\ \bibinfo {author} {\bibfnamefont {B.~D.}\ \bibnamefont {Gaulin}},\ }\bibfield  {title} {\bibinfo {title} {Octupolar versus n{\'e}el order in cubic 5d2 double perovskites},\ }\href {https://doi.org/10.1103/PhysRevLett.124.087206} {\bibfield  {journal} {\bibinfo  {journal} {Physical Review Letters}\ }\textbf {\bibinfo {volume} {124}},\ \bibinfo {pages} {87206} (\bibinfo {year} {2020})}\BibitemShut {NoStop}%
\bibitem [{\citenamefont {Pourovskii}\ \emph {et~al.}(2021)\citenamefont {Pourovskii}, \citenamefont {Mosca},\ and\ \citenamefont {Franchini}}]{Pourovskii2021}%
  \BibitemOpen
  \bibfield  {author} {\bibinfo {author} {\bibfnamefont {L.~V.}\ \bibnamefont {Pourovskii}}, \bibinfo {author} {\bibfnamefont {D.~F.}\ \bibnamefont {Mosca}},\ and\ \bibinfo {author} {\bibfnamefont {C.}~\bibnamefont {Franchini}},\ }\bibfield  {title} {\bibinfo {title} {Ferro-octupolar order and low-energy excitations in d2 double perovskites of osmium},\ }\href@noop {} {\bibfield  {journal} {\bibinfo  {journal} {Physical Review Letters}\ }\textbf {\bibinfo {volume} {127}},\ \bibinfo {pages} {237201} (\bibinfo {year} {2021})},\ \Eprint {https://arxiv.org/abs/2107.04493v1} {arXiv:2107.04493v1} \BibitemShut {NoStop}%
\bibitem [{\citenamefont {Takayama}\ \emph {et~al.}(2021)\citenamefont {Takayama}, \citenamefont {Chaloupka}, \citenamefont {Smerald}, \citenamefont {Khaliullin},\ and\ \citenamefont {Takagi}}]{Takayama2021}%
  \BibitemOpen
  \bibfield  {author} {\bibinfo {author} {\bibfnamefont {T.}~\bibnamefont {Takayama}}, \bibinfo {author} {\bibfnamefont {J.}~\bibnamefont {Chaloupka}}, \bibinfo {author} {\bibfnamefont {A.}~\bibnamefont {Smerald}}, \bibinfo {author} {\bibfnamefont {G.}~\bibnamefont {Khaliullin}},\ and\ \bibinfo {author} {\bibfnamefont {H.}~\bibnamefont {Takagi}},\ }\bibfield  {title} {\bibinfo {title} {Spin-orbit-entangled electronic phases in 4d and 5d transition-metal compounds},\ }\href {https://doi.org/10.7566/JPSJ.90.062001} {\bibfield  {journal} {\bibinfo  {journal} {Journal of the Physical Society of Japan}\ }\textbf {\bibinfo {volume} {90}},\ \bibinfo {pages} {062001} (\bibinfo {year} {2021})},\ \Eprint {https://arxiv.org/abs/2102.02740} {arXiv:2102.02740} \BibitemShut {NoStop}%
\bibitem [{\citenamefont {Anisimov}\ \emph {et~al.}(1999)\citenamefont {Anisimov}, \citenamefont {Bukhvalov},\ and\ \citenamefont {Rice}}]{anisimov1999}%
  \BibitemOpen
  \bibfield  {author} {\bibinfo {author} {\bibfnamefont {V.}~\bibnamefont {Anisimov}}, \bibinfo {author} {\bibfnamefont {D.}~\bibnamefont {Bukhvalov}},\ and\ \bibinfo {author} {\bibfnamefont {T.}~\bibnamefont {Rice}},\ }\bibfield  {title} {\bibinfo {title} {Electronic structure of possible nickelate analogs to the cuprates},\ }\href@noop {} {\bibfield  {journal} {\bibinfo  {journal} {Physical Review B}\ }\textbf {\bibinfo {volume} {59}},\ \bibinfo {pages} {7901} (\bibinfo {year} {1999})}\BibitemShut {NoStop}%
\bibitem [{\citenamefont {Li}\ \emph {et~al.}(2019)\citenamefont {Li}, \citenamefont {Lee}, \citenamefont {Wang}, \citenamefont {Osada}, \citenamefont {Crossley}, \citenamefont {Lee}, \citenamefont {Cui}, \citenamefont {Hikita},\ and\ \citenamefont {Hwang}}]{li2019}%
  \BibitemOpen
  \bibfield  {author} {\bibinfo {author} {\bibfnamefont {D.}~\bibnamefont {Li}}, \bibinfo {author} {\bibfnamefont {K.}~\bibnamefont {Lee}}, \bibinfo {author} {\bibfnamefont {B.~Y.}\ \bibnamefont {Wang}}, \bibinfo {author} {\bibfnamefont {M.}~\bibnamefont {Osada}}, \bibinfo {author} {\bibfnamefont {S.}~\bibnamefont {Crossley}}, \bibinfo {author} {\bibfnamefont {H.~R.}\ \bibnamefont {Lee}}, \bibinfo {author} {\bibfnamefont {Y.}~\bibnamefont {Cui}}, \bibinfo {author} {\bibfnamefont {Y.}~\bibnamefont {Hikita}},\ and\ \bibinfo {author} {\bibfnamefont {H.~Y.}\ \bibnamefont {Hwang}},\ }\bibfield  {title} {\bibinfo {title} {Superconductivity in an infinite-layer nickelate},\ }\href@noop {} {\bibfield  {journal} {\bibinfo  {journal} {Nature}\ }\textbf {\bibinfo {volume} {572}},\ \bibinfo {pages} {624} (\bibinfo {year} {2019})}\BibitemShut {NoStop}%
\bibitem [{\citenamefont {Hepting}\ \emph {et~al.}(2020)\citenamefont {Hepting}, \citenamefont {Li}, \citenamefont {Jia}, \citenamefont {Lu}, \citenamefont {Paris}, \citenamefont {Tseng}, \citenamefont {Feng}, \citenamefont {Osada}, \citenamefont {Been}, \citenamefont {Hikita} \emph {et~al.}}]{hepting2020}%
  \BibitemOpen
  \bibfield  {author} {\bibinfo {author} {\bibfnamefont {M.}~\bibnamefont {Hepting}}, \bibinfo {author} {\bibfnamefont {D.}~\bibnamefont {Li}}, \bibinfo {author} {\bibfnamefont {C.}~\bibnamefont {Jia}}, \bibinfo {author} {\bibfnamefont {H.}~\bibnamefont {Lu}}, \bibinfo {author} {\bibfnamefont {E.}~\bibnamefont {Paris}}, \bibinfo {author} {\bibfnamefont {Y.}~\bibnamefont {Tseng}}, \bibinfo {author} {\bibfnamefont {X.}~\bibnamefont {Feng}}, \bibinfo {author} {\bibfnamefont {M.}~\bibnamefont {Osada}}, \bibinfo {author} {\bibfnamefont {E.}~\bibnamefont {Been}}, \bibinfo {author} {\bibfnamefont {Y.}~\bibnamefont {Hikita}}, \emph {et~al.},\ }\bibfield  {title} {\bibinfo {title} {Electronic structure of the parent compound of superconducting infinite-layer nickelates},\ }\href@noop {} {\bibfield  {journal} {\bibinfo  {journal} {Nature materials}\ }\textbf {\bibinfo {volume} {19}},\ \bibinfo {pages} {381} (\bibinfo {year} {2020})}\BibitemShut {NoStop}%
\bibitem [{\citenamefont {Mitchell}(2015)}]{mitchell2015}%
  \BibitemOpen
  \bibfield  {author} {\bibinfo {author} {\bibfnamefont {J.}~\bibnamefont {Mitchell}},\ }\bibfield  {title} {\bibinfo {title} {Sr$_2${I}r{O}$_4$: Gateway to cuprate superconductivity?},\ }\href@noop {} {\bibfield  {journal} {\bibinfo  {journal} {APL Materials}\ }\textbf {\bibinfo {volume} {3}} (\bibinfo {year} {2015})}\BibitemShut {NoStop}%
\bibitem [{\citenamefont {Arita}\ \emph {et~al.}(2007)\citenamefont {Arita}, \citenamefont {Yamasaki}, \citenamefont {Held}, \citenamefont {Matsuno},\ and\ \citenamefont {Kuroki}}]{arita2007}%
  \BibitemOpen
  \bibfield  {author} {\bibinfo {author} {\bibfnamefont {R.}~\bibnamefont {Arita}}, \bibinfo {author} {\bibfnamefont {A.}~\bibnamefont {Yamasaki}}, \bibinfo {author} {\bibfnamefont {K.}~\bibnamefont {Held}}, \bibinfo {author} {\bibfnamefont {J.}~\bibnamefont {Matsuno}},\ and\ \bibinfo {author} {\bibfnamefont {K.}~\bibnamefont {Kuroki}},\ }\bibfield  {title} {\bibinfo {title} {Sr$_2${VO}$_4$ and {B}a$_2${VO}$_4$ under pressure: An orbital switch and potential d$^1$ superconductor},\ }\href {https://doi.org/10.1103/PhysRevB.75.174521} {\bibfield  {journal} {\bibinfo  {journal} {Physical Review B}\ }\textbf {\bibinfo {volume} {75}},\ \bibinfo {pages} {174521} (\bibinfo {year} {2007})}\BibitemShut {NoStop}%
\bibitem [{\citenamefont {Kasimov}\ \emph {et~al.}(1974)\citenamefont {Kasimov}, \citenamefont {Vovkotrub},\ and\ \citenamefont {Krylov}}]{Kasimov1974}%
  \BibitemOpen
  \bibfield  {author} {\bibinfo {author} {\bibfnamefont {G.}~\bibnamefont {Kasimov}}, \bibinfo {author} {\bibfnamefont {E.}~\bibnamefont {Vovkotrub}},\ and\ \bibinfo {author} {\bibfnamefont {E.}~\bibnamefont {Krylov}},\ }\bibfield  {title} {\bibinfo {title} {Synthesis of strontium orthoniobate},\ }\href@noop {} {\bibfield  {journal} {\bibinfo  {journal} {J. Inorg. Chem.}\ }\textbf {\bibinfo {volume} {19}},\ \bibinfo {pages} {148} (\bibinfo {year} {1974})}\BibitemShut {NoStop}%
\bibitem [{Vil()}]{Villars2023:sm_isp_sd_1502124}%
  \BibitemOpen
  \href {https://materials.springer.com/isp/crystallographic/docs/sd_1502124} {\bibinfo {title} {Sr$_2${T}i{O}$_4$ crystal structure: Datasheet from ``pauling file multinaries edition -- 2022'' in springermaterials}},\ \bibinfo {note} {copyright 2023 Springer-Verlag Berlin Heidelberg {\&} Material Phases Data System (MPDS), Switzerland {\&} National Institute for Materials Science (NIMS), Japan}\BibitemShut {NoStop}%
\bibitem [{\citenamefont {Isawa}\ and\ \citenamefont {Nagano}(2001)}]{Isawa2001}%
  \BibitemOpen
  \bibfield  {author} {\bibinfo {author} {\bibfnamefont {K.}~\bibnamefont {Isawa}}\ and\ \bibinfo {author} {\bibfnamefont {M.}~\bibnamefont {Nagano}},\ }\bibfield  {title} {\bibinfo {title} {Synthesis and physical properties of niobium-based oxide, $\mathrm{Sr_{2−x}La_{x}NbO_{4} (0 \leq x < 0.2)}$},\ }\href {https://doi.org/https://doi.org/10.1016/S0921-4534(01)00245-3} {\bibfield  {journal} {\bibinfo  {journal} {Physica C: Superconductivity}\ }\textbf {\bibinfo {volume} {357-360}},\ \bibinfo {pages} {359} (\bibinfo {year} {2001})}\BibitemShut {NoStop}%
\bibitem [{\citenamefont {Nakamura}(1994)}]{nakamura1994}%
  \BibitemOpen
  \bibfield  {author} {\bibinfo {author} {\bibfnamefont {A.}~\bibnamefont {Nakamura}},\ }\bibfield  {title} {\bibinfo {title} {{A}$_2${N}b$_{1+X}${O}$_y$({A}={C}a,{S}r)},\ }\href {https://doi.org/10.1143/JJAP.33.L583} {\bibfield  {journal} {\bibinfo  {journal} {Japanese Journal of Applied Physics}\ }\textbf {\bibinfo {volume} {33}},\ \bibinfo {pages} {L583} (\bibinfo {year} {1994})}\BibitemShut {NoStop}%
\bibitem [{\citenamefont {Ueno}\ \emph {et~al.}(2014)\citenamefont {Ueno}, \citenamefont {Kim}, \citenamefont {Takata},\ and\ \citenamefont {Katsufuji}}]{ueno2014}%
  \BibitemOpen
  \bibfield  {author} {\bibinfo {author} {\bibfnamefont {T.}~\bibnamefont {Ueno}}, \bibinfo {author} {\bibfnamefont {J.}~\bibnamefont {Kim}}, \bibinfo {author} {\bibfnamefont {M.}~\bibnamefont {Takata}},\ and\ \bibinfo {author} {\bibfnamefont {T.}~\bibnamefont {Katsufuji}},\ }\bibfield  {title} {\bibinfo {title} {Effect of offstoichiometry on the physical properties of sr2vo4},\ }\bibfield  {journal} {\bibinfo  {journal} {Journal of the Physical Society of Japan}\ }\textbf {\bibinfo {volume} {83}},\ \href {https://doi.org/10.7566/JPSJ.83.034708} {10.7566/JPSJ.83.034708} (\bibinfo {year} {2014})\BibitemShut {NoStop}%
\bibitem [{\citenamefont {Perdew}\ \emph {et~al.}(1997)\citenamefont {Perdew}, \citenamefont {Burke},\ and\ \citenamefont {Ernzerhof}}]{Perdew1997}%
  \BibitemOpen
  \bibfield  {author} {\bibinfo {author} {\bibfnamefont {J.~P.}\ \bibnamefont {Perdew}}, \bibinfo {author} {\bibfnamefont {K.}~\bibnamefont {Burke}},\ and\ \bibinfo {author} {\bibfnamefont {M.}~\bibnamefont {Ernzerhof}},\ }\bibfield  {title} {\bibinfo {title} {Generalized gradient approximation made simple},\ }\href {https://doi.org/10.1103/PhysRevLett.78.1396} {\bibfield  {journal} {\bibinfo  {journal} {Phys. Rev. Lett.}\ }\textbf {\bibinfo {volume} {78}},\ \bibinfo {pages} {1396} (\bibinfo {year} {1997})}\BibitemShut {NoStop}%
\bibitem [{\citenamefont {Kresse}\ and\ \citenamefont {Furthm\"uller}(1996)}]{Kresse1996}%
  \BibitemOpen
  \bibfield  {author} {\bibinfo {author} {\bibfnamefont {G.}~\bibnamefont {Kresse}}\ and\ \bibinfo {author} {\bibfnamefont {J.}~\bibnamefont {Furthm\"uller}},\ }\bibfield  {title} {\bibinfo {title} {Efficient iterative schemes for ab initio total-energy calculations using a plane-wave basis set},\ }\href {https://doi.org/10.1103/PhysRevB.54.11169} {\bibfield  {journal} {\bibinfo  {journal} {Phys. Rev. B}\ }\textbf {\bibinfo {volume} {54}},\ \bibinfo {pages} {11169} (\bibinfo {year} {1996})}\BibitemShut {NoStop}%
\bibitem [{\citenamefont {Monkhorst}\ and\ \citenamefont {Pack}(1976)}]{Monkhorst1976}%
  \BibitemOpen
  \bibfield  {author} {\bibinfo {author} {\bibfnamefont {H.~J.}\ \bibnamefont {Monkhorst}}\ and\ \bibinfo {author} {\bibfnamefont {J.~D.}\ \bibnamefont {Pack}},\ }\bibfield  {title} {\bibinfo {title} {Special points for brillouin-zone integrations},\ }\href {https://doi.org/10.1103/PhysRevB.13.5188} {\bibfield  {journal} {\bibinfo  {journal} {Phys. Rev. B}\ }\textbf {\bibinfo {volume} {13}},\ \bibinfo {pages} {5188} (\bibinfo {year} {1976})}\BibitemShut {NoStop}%
\bibitem [{\citenamefont {Press}\ \emph {et~al.}(1986)\citenamefont {Press}, \citenamefont {Flannery}, \citenamefont {Teukolsky},\ and\ \citenamefont {Vetterling}}]{Press1986}%
  \BibitemOpen
  \bibfield  {author} {\bibinfo {author} {\bibfnamefont {W.}~\bibnamefont {Press}}, \bibinfo {author} {\bibfnamefont {B.}~\bibnamefont {Flannery}}, \bibinfo {author} {\bibfnamefont {S.}~\bibnamefont {Teukolsky}},\ and\ \bibinfo {author} {\bibfnamefont {W.}~\bibnamefont {Vetterling}},\ }\href@noop {} {\emph {\bibinfo {title} {Numerical recipes : the art of scientific computing}}}\ (\bibinfo  {publisher} {Cambridge, New York, Cambridge University Press},\ \bibinfo {year} {1986})\BibitemShut {NoStop}%
\bibitem [{\citenamefont {Momma}\ and\ \citenamefont {Izumi}(2011)}]{VESTA}%
  \BibitemOpen
  \bibfield  {author} {\bibinfo {author} {\bibfnamefont {K.}~\bibnamefont {Momma}}\ and\ \bibinfo {author} {\bibfnamefont {F.}~\bibnamefont {Izumi}},\ }\bibfield  {title} {\bibinfo {title} {{\it VESTA 3} for three-dimensional visualization of crystal, volumetric and morphology data},\ }\href {https://doi.org/10.1107/S0021889811038970} {\bibfield  {journal} {\bibinfo  {journal} {J. Appl. Crystallography}\ }\textbf {\bibinfo {volume} {44}},\ \bibinfo {pages} {1272} (\bibinfo {year} {2011})}\BibitemShut {NoStop}%
\bibitem [{\citenamefont {Liechtenstein}\ \emph {et~al.}(1995)\citenamefont {Liechtenstein}, \citenamefont {Anisimov},\ and\ \citenamefont {Zaanen}}]{Liechtenstein1995}%
  \BibitemOpen
  \bibfield  {author} {\bibinfo {author} {\bibfnamefont {A.~I.}\ \bibnamefont {Liechtenstein}}, \bibinfo {author} {\bibfnamefont {V.~I.}\ \bibnamefont {Anisimov}},\ and\ \bibinfo {author} {\bibfnamefont {J.}~\bibnamefont {Zaanen}},\ }\bibfield  {title} {\bibinfo {title} {Density-functional theory and strong interactions: Orbital ordering in mott-hubbard insulators},\ }\href {https://doi.org/10.1103/PhysRevB.52.R5467} {\bibfield  {journal} {\bibinfo  {journal} {Phys. Rev. B}\ }\textbf {\bibinfo {volume} {52}},\ \bibinfo {pages} {R5467} (\bibinfo {year} {1995})}\BibitemShut {NoStop}%
\bibitem [{\citenamefont {Cococcioni}\ and\ \citenamefont {de~Gironcoli}(2005)}]{Cococcioni2005}%
  \BibitemOpen
  \bibfield  {author} {\bibinfo {author} {\bibfnamefont {M.}~\bibnamefont {Cococcioni}}\ and\ \bibinfo {author} {\bibfnamefont {S.}~\bibnamefont {de~Gironcoli}},\ }\bibfield  {title} {\bibinfo {title} {Linear response approach to the calculation of the effective interaction parameters in the $\mathrm{LDA}+\mathrm{U}$ method},\ }\href {https://doi.org/10.1103/PhysRevB.71.035105} {\bibfield  {journal} {\bibinfo  {journal} {Phys. Rev. B}\ }\textbf {\bibinfo {volume} {71}},\ \bibinfo {pages} {035105} (\bibinfo {year} {2005})}\BibitemShut {NoStop}%
\bibitem [{\citenamefont {\ifmmode \mbox{\c{S}}\else \c{S}\fi{}a\ifmmode \mbox{\c{s}}\else \c{s}\fi{}\ifmmode \imath \else \i \fi{}o\ifmmode~\breve{g}\else \u{g}\fi{}lu}\ \emph {et~al.}(2011)\citenamefont {\ifmmode \mbox{\c{S}}\else \c{S}\fi{}a\ifmmode \mbox{\c{s}}\else \c{s}\fi{}\ifmmode \imath \else \i \fi{}o\ifmmode~\breve{g}\else \u{g}\fi{}lu}, \citenamefont {Friedrich},\ and\ \citenamefont {Bl\"ugel}}]{Sasioglu2011}%
  \BibitemOpen
  \bibfield  {author} {\bibinfo {author} {\bibfnamefont {E.}~\bibnamefont {\ifmmode \mbox{\c{S}}\else \c{S}\fi{}a\ifmmode \mbox{\c{s}}\else \c{s}\fi{}\ifmmode \imath \else \i \fi{}o\ifmmode~\breve{g}\else \u{g}\fi{}lu}}, \bibinfo {author} {\bibfnamefont {C.}~\bibnamefont {Friedrich}},\ and\ \bibinfo {author} {\bibfnamefont {S.}~\bibnamefont {Bl\"ugel}},\ }\bibfield  {title} {\bibinfo {title} {Effective coulomb interaction in transition metals from constrained random-phase approximation},\ }\href {https://doi.org/10.1103/PhysRevB.83.121101} {\bibfield  {journal} {\bibinfo  {journal} {Phys. Rev. B}\ }\textbf {\bibinfo {volume} {83}},\ \bibinfo {pages} {121101} (\bibinfo {year} {2011})}\BibitemShut {NoStop}%
\bibitem [{\citenamefont {Poteryaev}\ \emph {et~al.}()\citenamefont {Poteryaev}, \citenamefont {Belozerov}, \citenamefont {Dyachenko}, \citenamefont {Korotin}, \citenamefont {Korotin}, \citenamefont {Shorikov}, \citenamefont {Skorikov}, \citenamefont {Skornyakov},\ and\ \citenamefont {Streltsov}}]{AMULET}%
  \BibitemOpen
  \bibfield  {author} {\bibinfo {author} {\bibfnamefont {A.}~\bibnamefont {Poteryaev}}, \bibinfo {author} {\bibfnamefont {A.}~\bibnamefont {Belozerov}}, \bibinfo {author} {\bibfnamefont {A.}~\bibnamefont {Dyachenko}}, \bibinfo {author} {\bibfnamefont {D.}~\bibnamefont {Korotin}}, \bibinfo {author} {\bibfnamefont {M.}~\bibnamefont {Korotin}}, \bibinfo {author} {\bibfnamefont {A.}~\bibnamefont {Shorikov}}, \bibinfo {author} {\bibfnamefont {N.}~\bibnamefont {Skorikov}}, \bibinfo {author} {\bibfnamefont {S.}~\bibnamefont {Skornyakov}},\ and\ \bibinfo {author} {\bibfnamefont {S.}~\bibnamefont {Streltsov}},\ }\href@noop {} {\bibinfo {title} {Amulet}}\BibitemShut {NoStop}%
\bibitem [{\citenamefont {Gull}\ \emph {et~al.}(2011)\citenamefont {Gull}, \citenamefont {Millis}, \citenamefont {Lichtenstein}, \citenamefont {Rubtsov}, \citenamefont {Troyer},\ and\ \citenamefont {Werner}}]{gull2011}%
  \BibitemOpen
  \bibfield  {author} {\bibinfo {author} {\bibfnamefont {E.}~\bibnamefont {Gull}}, \bibinfo {author} {\bibfnamefont {A.~J.}\ \bibnamefont {Millis}}, \bibinfo {author} {\bibfnamefont {A.~I.}\ \bibnamefont {Lichtenstein}}, \bibinfo {author} {\bibfnamefont {A.~N.}\ \bibnamefont {Rubtsov}}, \bibinfo {author} {\bibfnamefont {M.}~\bibnamefont {Troyer}},\ and\ \bibinfo {author} {\bibfnamefont {P.}~\bibnamefont {Werner}},\ }\bibfield  {title} {\bibinfo {title} {Continuous-time monte carlo methods for quantum impurity models},\ }\href@noop {} {\bibfield  {journal} {\bibinfo  {journal} {Reviews of Modern Physics}\ }\textbf {\bibinfo {volume} {83}},\ \bibinfo {pages} {349} (\bibinfo {year} {2011})}\BibitemShut {NoStop}%
\bibitem [{\citenamefont {Marzari}\ and\ \citenamefont {Vanderbilt}(1997)}]{Marzari1997}%
  \BibitemOpen
  \bibfield  {author} {\bibinfo {author} {\bibfnamefont {N.}~\bibnamefont {Marzari}}\ and\ \bibinfo {author} {\bibfnamefont {D.}~\bibnamefont {Vanderbilt}},\ }\bibfield  {title} {\bibinfo {title} {Maximally localized generalized wannier functions for composite energy bands},\ }\href {https://doi.org/10.1103/PhysRevB.56.12847} {\bibfield  {journal} {\bibinfo  {journal} {Phys. Rev. B}\ }\textbf {\bibinfo {volume} {56}},\ \bibinfo {pages} {12847} (\bibinfo {year} {1997})}\BibitemShut {NoStop}%
\bibitem [{\citenamefont {Marzari}\ \emph {et~al.}(2012)\citenamefont {Marzari}, \citenamefont {Mostofi}, \citenamefont {Yates}, \citenamefont {Souza},\ and\ \citenamefont {Vanderbilt}}]{Marzari2012}%
  \BibitemOpen
  \bibfield  {author} {\bibinfo {author} {\bibfnamefont {N.}~\bibnamefont {Marzari}}, \bibinfo {author} {\bibfnamefont {A.~A.}\ \bibnamefont {Mostofi}}, \bibinfo {author} {\bibfnamefont {J.~R.}\ \bibnamefont {Yates}}, \bibinfo {author} {\bibfnamefont {I.}~\bibnamefont {Souza}},\ and\ \bibinfo {author} {\bibfnamefont {D.}~\bibnamefont {Vanderbilt}},\ }\bibfield  {title} {\bibinfo {title} {Maximally localized wannier functions: Theory and applications},\ }\href {https://doi.org/10.1103/RevModPhys.84.1419} {\bibfield  {journal} {\bibinfo  {journal} {Rev. Mod. Phys.}\ }\textbf {\bibinfo {volume} {84}},\ \bibinfo {pages} {1419} (\bibinfo {year} {2012})}\BibitemShut {NoStop}%
\bibitem [{\citenamefont {Mostofi}\ \emph {et~al.}(2014)\citenamefont {Mostofi}, \citenamefont {Yates}, \citenamefont {Pizzi}, \citenamefont {Lee}, \citenamefont {Souza}, \citenamefont {Vanderbilt},\ and\ \citenamefont {Marzari}}]{Mostofi2014}%
  \BibitemOpen
  \bibfield  {author} {\bibinfo {author} {\bibfnamefont {A.~A.}\ \bibnamefont {Mostofi}}, \bibinfo {author} {\bibfnamefont {J.~R.}\ \bibnamefont {Yates}}, \bibinfo {author} {\bibfnamefont {G.}~\bibnamefont {Pizzi}}, \bibinfo {author} {\bibfnamefont {Y.~S.}\ \bibnamefont {Lee}}, \bibinfo {author} {\bibfnamefont {I.}~\bibnamefont {Souza}}, \bibinfo {author} {\bibfnamefont {D.}~\bibnamefont {Vanderbilt}},\ and\ \bibinfo {author} {\bibfnamefont {N.}~\bibnamefont {Marzari}},\ }\bibfield  {title} {\bibinfo {title} {An updated version of {W}annier90: A tool for obtaining maximally-localised wannier functions},\ }\href {https://doi.org/10.1016/j.cpc.2014.05.003} {\bibfield  {journal} {\bibinfo  {journal} {Computer Physics Communications}\ }\textbf {\bibinfo {volume} {185}},\ \bibinfo {pages} {2309} (\bibinfo {year} {2014})}\BibitemShut {NoStop}%
\bibitem [{\citenamefont {Georges}\ \emph {et~al.}(2013)\citenamefont {Georges}, \citenamefont {Medici},\ and\ \citenamefont {Mravlje}}]{Georges2013}%
  \BibitemOpen
  \bibfield  {author} {\bibinfo {author} {\bibfnamefont {A.}~\bibnamefont {Georges}}, \bibinfo {author} {\bibfnamefont {L.~D.}\ \bibnamefont {Medici}},\ and\ \bibinfo {author} {\bibfnamefont {J.}~\bibnamefont {Mravlje}},\ }\bibfield  {title} {\bibinfo {title} {Strong correlations from hund's coupling},\ }\href {https://doi.org/10.1146/annurev-conmatphys-020911-125045} {\bibfield  {journal} {\bibinfo  {journal} {Annual Review of Condensed Matter Physics}\ }\textbf {\bibinfo {volume} {4}},\ \bibinfo {pages} {137} (\bibinfo {year} {2013})}\BibitemShut {NoStop}%
\bibitem [{Tar()}]{Taran2025}%
  \BibitemOpen
  \href@noop {} {}\bibinfo {note} {{See Supplemental material at [URL will be inserted by publisher] for details of crystal structure, energy-volume curves, total and projected DOSes}}\BibitemShut {NoStop}%
\bibitem [{\citenamefont {Ruddlesden}\ and\ \citenamefont {Popper}(1957)}]{Ruddlesden1957}%
  \BibitemOpen
  \bibfield  {author} {\bibinfo {author} {\bibfnamefont {S.~N.}\ \bibnamefont {Ruddlesden}}\ and\ \bibinfo {author} {\bibfnamefont {P.}~\bibnamefont {Popper}},\ }\bibfield  {title} {\bibinfo {title} {{New compounds of the {K}$_2${N}i{F}$_4$ type}},\ }\href {https://doi.org/10.1107/S0365110X57001929} {\bibfield  {journal} {\bibinfo  {journal} {Acta Crystallographica}\ }\textbf {\bibinfo {volume} {10}},\ \bibinfo {pages} {538} (\bibinfo {year} {1957})}\BibitemShut {NoStop}%
\bibitem [{\citenamefont {Ruddlesden}\ and\ \citenamefont {Popper}(1958)}]{Ruddlesden1958}%
  \BibitemOpen
  \bibfield  {author} {\bibinfo {author} {\bibfnamefont {S.~N.}\ \bibnamefont {Ruddlesden}}\ and\ \bibinfo {author} {\bibfnamefont {P.}~\bibnamefont {Popper}},\ }\bibfield  {title} {\bibinfo {title} {{The compound Sr${\sb 3}$Ti${\sb 2}$O${\sb 7}$ and its structure}},\ }\href {https://doi.org/10.1107/S0365110X58000128} {\bibfield  {journal} {\bibinfo  {journal} {Acta Crystallographica}\ }\textbf {\bibinfo {volume} {11}},\ \bibinfo {pages} {54} (\bibinfo {year} {1958})}\BibitemShut {NoStop}%
\bibitem [{\citenamefont {Wells}(1975)}]{Wells1975}%
  \BibitemOpen
  \bibfield  {author} {\bibinfo {author} {\bibfnamefont {A.}~\bibnamefont {Wells}},\ }\href@noop {} {\emph {\bibinfo {title} {Structural Inorganic Chemistry, 4th Edition}}}\ (\bibinfo  {publisher} {Oxford University Press},\ \bibinfo {year} {1975})\ p.\ \bibinfo {pages} {498}\BibitemShut {NoStop}%
\bibitem [{\citenamefont {Birch}(1947)}]{Birch1947}%
  \BibitemOpen
  \bibfield  {author} {\bibinfo {author} {\bibfnamefont {F.}~\bibnamefont {Birch}},\ }\bibfield  {title} {\bibinfo {title} {Finite elastic strain of cubic crystals},\ }\href {https://doi.org/10.1103/PhysRev.71.809} {\bibfield  {journal} {\bibinfo  {journal} {Phys. Rev.}\ }\textbf {\bibinfo {volume} {71}},\ \bibinfo {pages} {809} (\bibinfo {year} {1947})}\BibitemShut {NoStop}%
\bibitem [{\citenamefont {Mounet}\ \emph {et~al.}(2018)\citenamefont {Mounet}, \citenamefont {Gibertini}, \citenamefont {Schwaller}, \citenamefont {Campi}, \citenamefont {Merkys}, \citenamefont {Marrazzo}, \citenamefont {Sohier}, \citenamefont {Castelli}, \citenamefont {Cepellotti}, \citenamefont {Pizzi},\ and\ \citenamefont {Marzari}}]{Mounet2018}%
  \BibitemOpen
  \bibfield  {author} {\bibinfo {author} {\bibfnamefont {N.}~\bibnamefont {Mounet}}, \bibinfo {author} {\bibfnamefont {M.}~\bibnamefont {Gibertini}}, \bibinfo {author} {\bibfnamefont {P.}~\bibnamefont {Schwaller}}, \bibinfo {author} {\bibfnamefont {D.}~\bibnamefont {Campi}}, \bibinfo {author} {\bibfnamefont {A.}~\bibnamefont {Merkys}}, \bibinfo {author} {\bibfnamefont {A.}~\bibnamefont {Marrazzo}}, \bibinfo {author} {\bibfnamefont {T.}~\bibnamefont {Sohier}}, \bibinfo {author} {\bibfnamefont {I.}~\bibnamefont {Castelli}}, \bibinfo {author} {\bibfnamefont {A.}~\bibnamefont {Cepellotti}}, \bibinfo {author} {\bibfnamefont {G.}~\bibnamefont {Pizzi}},\ and\ \bibinfo {author} {\bibfnamefont {N.}~\bibnamefont {Marzari}},\ }\bibfield  {title} {\bibinfo {title} {Novel two-dimensional materials from high-throughput computational exfoliation of experimentally known compounds},\ }\href {https://doi.org/10.1038/s41565-017-0035-5} {\bibfield  {journal} {\bibinfo  {journal} {Nature Nanotechnology}\ }\textbf {\bibinfo
  {volume} {13}} (\bibinfo {year} {2018})}\BibitemShut {NoStop}%
\bibitem [{\citenamefont {Matsuno}\ \emph {et~al.}(2005)\citenamefont {Matsuno}, \citenamefont {Okimoto}, \citenamefont {Kawasaki},\ and\ \citenamefont {Tokura}}]{matsuno2005a}%
  \BibitemOpen
  \bibfield  {author} {\bibinfo {author} {\bibfnamefont {J.}~\bibnamefont {Matsuno}}, \bibinfo {author} {\bibfnamefont {Y.}~\bibnamefont {Okimoto}}, \bibinfo {author} {\bibfnamefont {M.}~\bibnamefont {Kawasaki}},\ and\ \bibinfo {author} {\bibfnamefont {Y.}~\bibnamefont {Tokura}},\ }\bibfield  {title} {\bibinfo {title} {Variation of the electronic structure in systematically synthesized sr2mo4 ( m = ti , v, cr, mn, and co)},\ }\href {https://doi.org/10.1103/PhysRevLett.95.176404} {\bibfield  {journal} {\bibinfo  {journal} {Physical Review Letters}\ }\textbf {\bibinfo {volume} {95}},\ \bibinfo {pages} {176404} (\bibinfo {year} {2005})}\BibitemShut {NoStop}%
\bibitem [{\citenamefont {Eremin}\ \emph {et~al.}(2011)\citenamefont {Eremin}, \citenamefont {Deisenhofer}, \citenamefont {Eremina}, \citenamefont {Teyssier}, \citenamefont {Van Der~Marel},\ and\ \citenamefont {Loidl}}]{Eremin2011}%
  \BibitemOpen
  \bibfield  {author} {\bibinfo {author} {\bibfnamefont {M.~V.}\ \bibnamefont {Eremin}}, \bibinfo {author} {\bibfnamefont {J.}~\bibnamefont {Deisenhofer}}, \bibinfo {author} {\bibfnamefont {R.~M.}\ \bibnamefont {Eremina}}, \bibinfo {author} {\bibfnamefont {J.}~\bibnamefont {Teyssier}}, \bibinfo {author} {\bibfnamefont {D.}~\bibnamefont {Van Der~Marel}},\ and\ \bibinfo {author} {\bibfnamefont {A.}~\bibnamefont {Loidl}},\ }\bibfield  {title} {\bibinfo {title} {Alternating spin-orbital order in tetragonal sr2vo4},\ }\href {https://doi.org/10.1103/PhysRevB.84.212407} {\bibfield  {journal} {\bibinfo  {journal} {Physical Review B}\ }\textbf {\bibinfo {volume} {84}},\ \bibinfo {pages} {4} (\bibinfo {year} {2011})}\BibitemShut {NoStop}%
\bibitem [{\citenamefont {Hybertsen}\ \emph {et~al.}(1990)\citenamefont {Hybertsen}, \citenamefont {Stechel}, \citenamefont {Schluter},\ and\ \citenamefont {Jennison}}]{hybertsen1990}%
  \BibitemOpen
  \bibfield  {author} {\bibinfo {author} {\bibfnamefont {M.~S.}\ \bibnamefont {Hybertsen}}, \bibinfo {author} {\bibfnamefont {E.}~\bibnamefont {Stechel}}, \bibinfo {author} {\bibfnamefont {M.}~\bibnamefont {Schluter}},\ and\ \bibinfo {author} {\bibfnamefont {D.}~\bibnamefont {Jennison}},\ }\bibfield  {title} {\bibinfo {title} {Renormalization from density-functional theory to strong-coupling models for electronic states in {C}u-{O} materials},\ }\href@noop {} {\bibfield  {journal} {\bibinfo  {journal} {Physical Review B}\ }\textbf {\bibinfo {volume} {41}},\ \bibinfo {pages} {11068} (\bibinfo {year} {1990})}\BibitemShut {NoStop}%
\bibitem [{\citenamefont {Abragam}\ and\ \citenamefont {Bleaney}(1970)}]{Abragam}%
  \BibitemOpen
  \bibfield  {author} {\bibinfo {author} {\bibfnamefont {A.}~\bibnamefont {Abragam}}\ and\ \bibinfo {author} {\bibfnamefont {B.}~\bibnamefont {Bleaney}},\ }\href@noop {} {\emph {\bibinfo {title} {Electron Paramagnetic Resonance of Transition Ions}}}\ (\bibinfo  {publisher} {Clarendon press},\ \bibinfo {address} {Oxford},\ \bibinfo {year} {1970})\BibitemShut {NoStop}%
\bibitem [{\citenamefont {Streltsov}(2013)}]{streltsov2013}%
  \BibitemOpen
  \bibfield  {author} {\bibinfo {author} {\bibfnamefont {S.}~\bibnamefont {Streltsov}},\ }\bibfield  {title} {\bibinfo {title} {Magnetic moment suppression in {B}a$_3${C}o{R}u$_2${O}$_9$: Hybridization effect},\ }\href {https://doi.org/10.1103/PhysRevB.88.024429} {\bibfield  {journal} {\bibinfo  {journal} {Physical Review B}\ }\textbf {\bibinfo {volume} {88}},\ \bibinfo {pages} {024429} (\bibinfo {year} {2013})}\BibitemShut {NoStop}%
\bibitem [{\citenamefont {Katanin}\ \emph {et~al.}(2023)\citenamefont {Katanin}, \citenamefont {Belozerov}, \citenamefont {Lichtenstein},\ and\ \citenamefont {Katsnelson}}]{katanin2023}%
  \BibitemOpen
  \bibfield  {author} {\bibinfo {author} {\bibfnamefont {A.}~\bibnamefont {Katanin}}, \bibinfo {author} {\bibfnamefont {A.}~\bibnamefont {Belozerov}}, \bibinfo {author} {\bibfnamefont {A.}~\bibnamefont {Lichtenstein}},\ and\ \bibinfo {author} {\bibfnamefont {M.}~\bibnamefont {Katsnelson}},\ }\bibfield  {title} {\bibinfo {title} {Exchange interactions in iron and nickel: Dft+ dmft study in paramagnetic phase},\ }\href@noop {} {\bibfield  {journal} {\bibinfo  {journal} {Physical Review B}\ }\textbf {\bibinfo {volume} {107}},\ \bibinfo {pages} {235118} (\bibinfo {year} {2023})}\BibitemShut {NoStop}%
\bibitem [{\citenamefont {Igoshev}\ \emph {et~al.}(2013)\citenamefont {Igoshev}, \citenamefont {Efremov}, \citenamefont {Poteryaev}, \citenamefont {Katanin},\ and\ \citenamefont {Anisimov}}]{igoshev2013}%
  \BibitemOpen
  \bibfield  {author} {\bibinfo {author} {\bibfnamefont {P.}~\bibnamefont {Igoshev}}, \bibinfo {author} {\bibfnamefont {A.}~\bibnamefont {Efremov}}, \bibinfo {author} {\bibfnamefont {A.}~\bibnamefont {Poteryaev}}, \bibinfo {author} {\bibfnamefont {A.}~\bibnamefont {Katanin}},\ and\ \bibinfo {author} {\bibfnamefont {V.}~\bibnamefont {Anisimov}},\ }\bibfield  {title} {\bibinfo {title} {Magnetic fluctuations and effective magnetic moments in $\gamma$-iron due to electronic structure peculiarities},\ }\href@noop {} {\bibfield  {journal} {\bibinfo  {journal} {Physical Review B—Condensed Matter and Materials Physics}\ }\textbf {\bibinfo {volume} {88}},\ \bibinfo {pages} {155120} (\bibinfo {year} {2013})}\BibitemShut {NoStop}%
\bibitem [{\citenamefont {Streltsov}\ \emph {et~al.}(2025)\citenamefont {Streltsov}, \citenamefont {Takegami}, \citenamefont {Nakamura}, \citenamefont {Kovaleva}, \citenamefont {Poteryaev}, \citenamefont {Nikolaev}, \citenamefont {Xu}, \citenamefont {Sui}, \citenamefont {Yoshimura}, \citenamefont {Tsuei}, \citenamefont {Saini}, \citenamefont {Khomskii},\ and\ \citenamefont {Mizokawa}}]{streltsov2025}%
  \BibitemOpen
  \bibfield  {author} {\bibinfo {author} {\bibfnamefont {S.~V.}\ \bibnamefont {Streltsov}}, \bibinfo {author} {\bibfnamefont {D.}~\bibnamefont {Takegami}}, \bibinfo {author} {\bibfnamefont {R.}~\bibnamefont {Nakamura}}, \bibinfo {author} {\bibfnamefont {P.~P.}\ \bibnamefont {Kovaleva}}, \bibinfo {author} {\bibfnamefont {A.~I.}\ \bibnamefont {Poteryaev}}, \bibinfo {author} {\bibfnamefont {S.~A.}\ \bibnamefont {Nikolaev}}, \bibinfo {author} {\bibfnamefont {H.-H.}\ \bibnamefont {Xu}}, \bibinfo {author} {\bibfnamefont {Y.}~\bibnamefont {Sui}}, \bibinfo {author} {\bibfnamefont {M.}~\bibnamefont {Yoshimura}}, \bibinfo {author} {\bibfnamefont {K.-D.}\ \bibnamefont {Tsuei}}, \bibinfo {author} {\bibfnamefont {N.~L.}\ \bibnamefont {Saini}}, \bibinfo {author} {\bibfnamefont {D.~I.}\ \bibnamefont {Khomskii}},\ and\ \bibinfo {author} {\bibfnamefont {T.}~\bibnamefont {Mizokawa}},\ }\bibfield  {title} {\bibinfo {title} {Beyond a cluster-mott state in the breathing kagome lattice of lizn2mo3o8},\ }\href
  {https://doi.org/10.1103/PhysRevB.111.085124} {\bibfield  {journal} {\bibinfo  {journal} {Physical Review B}\ }\textbf {\bibinfo {volume} {111}},\ \bibinfo {pages} {085124} (\bibinfo {year} {2025})}\BibitemShut {NoStop}%
\bibitem [{\citenamefont {Moriya}(2012)}]{moriya2012}%
  \BibitemOpen
  \bibfield  {author} {\bibinfo {author} {\bibfnamefont {T.}~\bibnamefont {Moriya}},\ }\href@noop {} {\emph {\bibinfo {title} {Spin Fluctuations in Itinerant Electron Magnetism}}}\ (\bibinfo  {publisher} {Springer Berlin Heidelberg},\ \bibinfo {year} {2012})\BibitemShut {NoStop}%
\bibitem [{\citenamefont {Vaugier}\ \emph {et~al.}(2012)\citenamefont {Vaugier}, \citenamefont {Jiang},\ and\ \citenamefont {Biermann}}]{Vaugier2012}%
  \BibitemOpen
  \bibfield  {author} {\bibinfo {author} {\bibfnamefont {L.}~\bibnamefont {Vaugier}}, \bibinfo {author} {\bibfnamefont {H.}~\bibnamefont {Jiang}},\ and\ \bibinfo {author} {\bibfnamefont {S.}~\bibnamefont {Biermann}},\ }\bibfield  {title} {\bibinfo {title} {Hubbard u and hund exchange j in transition metal oxides: Screening versus localization trends from constrained random phase approximation},\ }\href {https://doi.org/10.1103/PhysRevB.86.165105} {\bibfield  {journal} {\bibinfo  {journal} {Phys. Rev. B}\ }\textbf {\bibinfo {volume} {86}},\ \bibinfo {pages} {165105} (\bibinfo {year} {2012})}\BibitemShut {NoStop}%
\bibitem [{\citenamefont {Moore}\ \emph {et~al.}(2024)\citenamefont {Moore}, \citenamefont {Horton}, \citenamefont {Linscott}, \citenamefont {Ganose}, \citenamefont {Siron}, \citenamefont {O'Regan},\ and\ \citenamefont {Persson}}]{moore2024}%
  \BibitemOpen
  \bibfield  {author} {\bibinfo {author} {\bibfnamefont {G.~C.}\ \bibnamefont {Moore}}, \bibinfo {author} {\bibfnamefont {M.~K.}\ \bibnamefont {Horton}}, \bibinfo {author} {\bibfnamefont {E.}~\bibnamefont {Linscott}}, \bibinfo {author} {\bibfnamefont {A.~M.}\ \bibnamefont {Ganose}}, \bibinfo {author} {\bibfnamefont {M.}~\bibnamefont {Siron}}, \bibinfo {author} {\bibfnamefont {D.~D.}\ \bibnamefont {O'Regan}},\ and\ \bibinfo {author} {\bibfnamefont {K.~A.}\ \bibnamefont {Persson}},\ }\bibfield  {title} {\bibinfo {title} {High-throughput determination of hubbard u and hund j values for transition metal oxides via the linear response formalism},\ }\href {https://doi.org/10.1103/PhysRevMaterials.8.014409} {\bibfield  {journal} {\bibinfo  {journal} {Physical Review Materials}\ }\textbf {\bibinfo {volume} {8}},\ \bibinfo {pages} {014409} (\bibinfo {year} {2024})}\BibitemShut {NoStop}%
\bibitem [{\citenamefont {Paul}\ and\ \citenamefont {Birol}(2019)}]{paul2019}%
  \BibitemOpen
  \bibfield  {author} {\bibinfo {author} {\bibfnamefont {A.}~\bibnamefont {Paul}}\ and\ \bibinfo {author} {\bibfnamefont {T.}~\bibnamefont {Birol}},\ }\bibfield  {title} {\bibinfo {title} {Strain tuning of plasma frequency in vanadate, niobate, and molybdate perovskite oxides},\ }\href {https://doi.org/10.1103/PhysRevMaterials.3.085001} {\bibfield  {journal} {\bibinfo  {journal} {Physical Review Materials}\ }\textbf {\bibinfo {volume} {3}},\ \bibinfo {pages} {085001} (\bibinfo {year} {2019})}\BibitemShut {NoStop}%
\end{thebibliography}%
\

\end{document}


\title{Supplemental material for: Sr$_2$NbO$_4$: A $4d$ analogue of the layered perovskite Sr$_2$VO$_4$}

\author{Leonid S. Taran}
\email{leonidtaran97@gmail.com}
\affiliation{M. N. Mikheev Institute of Metal Physics, Ural Branch of Russian Academy of Sciences, 620137 Yekaterinburg, Russia}

\author{Anastasia E. Lebedeva}
\affiliation{Institute of Physics and Technology, Ural Federal University, 620002 Yekaterinburg, Russia}

\author{Sergey V. Streltsov}
\affiliation{M. N. Mikheev Institute of Metal Physics, Ural Branch of Russian Academy of Sciences, 620137 Yekaterinburg, Russia}
\affiliation{Institute of Physics and Technology, Ural Federal University, 620002 Yekaterinburg, Russia}

\date{\today}

\maketitle

\section{Crystal structure details}
Table. \ref{table:struct} presents the lattice parameters and interatomic distances for the Sr$_2$NbO$_4$ crystal structure obtained by relaxation. Atomic coordinates  are shown in Table \ref{table:AtomCoords}.

\begin{table}[h!]
    \centering
    \caption{Conventional unit cell parameters and interatomic distances for the Sr$_2$NbO$_4$ after relaxation using DFT method. Space group is $I4/mmm$.}
    \begin{ruledtabular}
    \begin{tabular}{lc}
        Parameter    &  Value \\ 
        \midrule 
        $a=b$ (\AA) & 4.03869 \\ 
        $c$ (\AA) & 12.75256  \\ 
        $\alpha$ (deg) & 90 \\ 
        $\beta$ (deg)  & 90 \\ 
        $\gamma$ (deg) & 90 \\ 
        $V$ (\AA$^3$)  & 208.00723 \\ 
        \midrule
        \multicolumn{2}{c}{Interatomic distances (\AA)} \\
        \midrule 
        M-O1 ($\times 4$)   & 2.01935 \\ 
        M-O2 ($\times 2$)   & 2.07834 \\ 
        Sr-M ($\times 4$)   & 3.41071 \\ 
        Sr-O1 ($\times 4$)  & 2.74867 \\
        Sr-O2 ($\times 4$)  & 2.86376 \\
        Sr-O2               & 2.43316 \\
        O1-O1 ($\times 4$)	& 2.85579 \\
        O1-O2 ($\times 4$)	& 2.89780 \\
    \end{tabular}
    \end{ruledtabular}
    \label{table:struct}
\end{table}

\begin{table}[h!]
    \centering
    \caption{Atomic coordinates for Sr$_2$NbO$_4$ conventional unit cell as obtained by the relaxation of the crystal structure in DFT. Space group is $I4/mmm$.}
    \begin{ruledtabular}
    \begin{tabular}{lccc}
        Site & $x$ & $y$ & $z$ \\
        \cmidrule(lr){1-4}
        \,Sr (4\textit{e}) & 0 & 0 & 0.35377  \\ 
        \,Nb (2\textit{a}) & 0 & 0 & 0 \\ 
        \,O1 (4\textit{c}) & 0 & 0.5 & 0 \\ 
        \,O2 (4\textit{e}) & 0 & 0 & 0.16297 \\ 
    \end{tabular}
    \end{ruledtabular}
    \label{table:AtomCoords}
\end{table}

\begin{figure}[h!]
\centering
\includegraphics[width=1\columnwidth]{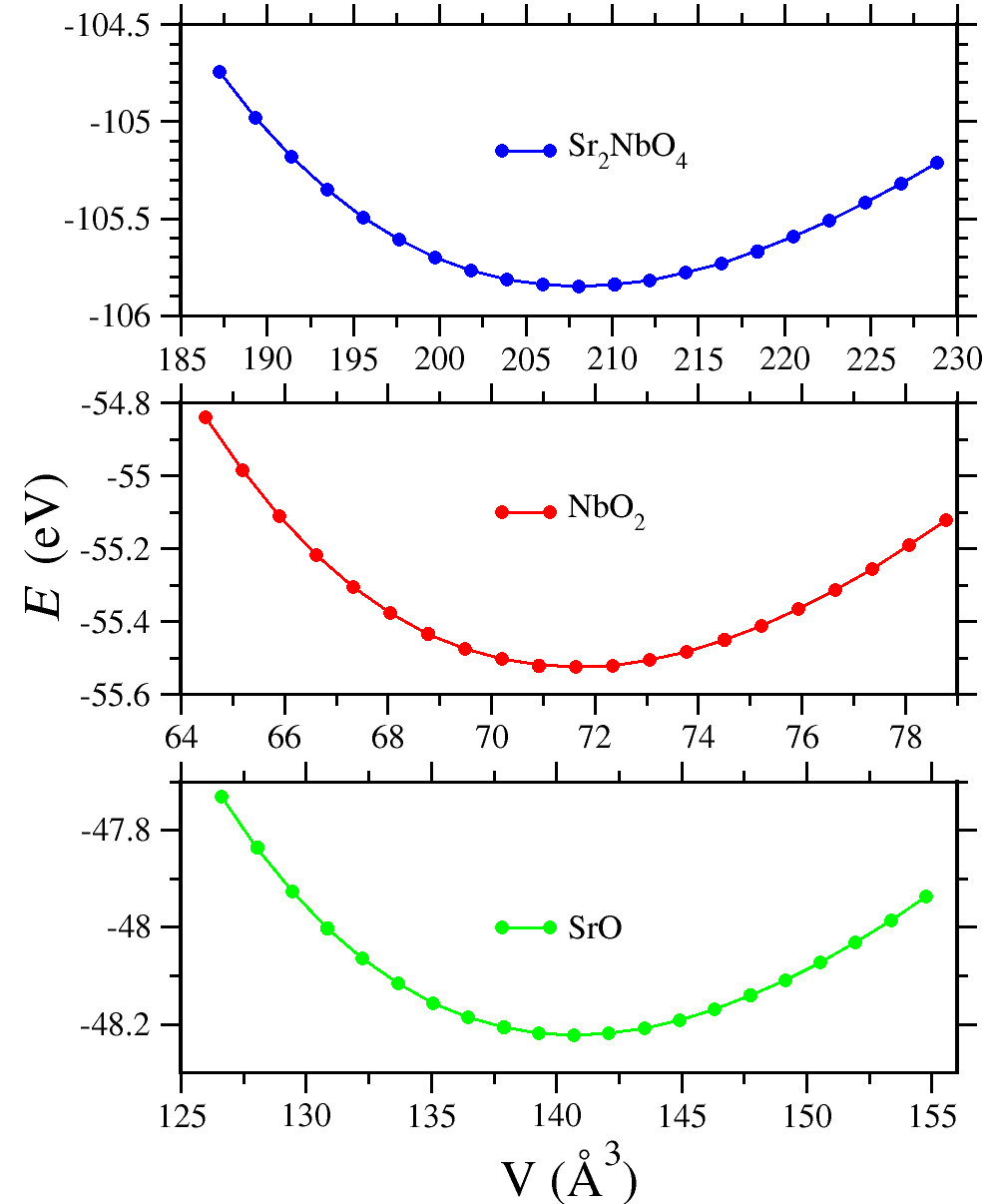}
    \caption{The energy-volume curves for Sr$_2$NbO$_4$, NbO$_2$ and SrO from DFT calculations.}
    \label{Fig: EV curves}
\end{figure}

The curves presented in Fig. \ref{Fig: EV curves} show the energy dependence on the volume of the structures involved in the formation reaction. The smoothness of the curves is due to the absence of phase transition in the considered pressure interval in Fig. 2 of the main paper. The minimum energy corresponds to the structure with the equilibrium volume at atmospheric pressure.

Table \ref{table:ICOHP} shows the negative values of ICOHP for the corresponding bonds from Table \ref{table:struct}. As can be seen, strontium is much weaker bonded with oxygen than niobium. Taking into account that the unit cell contains only two Sr-O bonds along the $ab$ plane, it is their breaking that is the most favorable for cleavage.

\begin{table}[h!]
    \centering
    \caption{Average -ICOHP results for Sr$_2$NbO$_4$ bondings.}
    \begin{ruledtabular}
    \begin{tabular}{lc}
      Bond & -ICOHP \\ \midrule
        Nb-O1 ($\times 4$)   & 4.46 \\ 
        Nb-O2 ($\times 2$)   & 4.04 \\
        Sr-Nb ($\times 4$)  & 0.45 \\
        Sr-O1 ($\times 4$)  & 0.23 \\
        Sr-O2 ($\times 4$)  & 0.42 \\
        Sr-O2               & 0.71 \\
        O1-O1 ($\times 4$)  & 0.10 \\
        O1-O2 ($\times 2$)  & 0.12 \\
    \end{tabular}
    \end{ruledtabular}
    \label{table:ICOHP}
\end{table}

\section{Non-magnetic DFT results}

\begin{figure}[h]
\centering
\includegraphics[width=1\columnwidth]{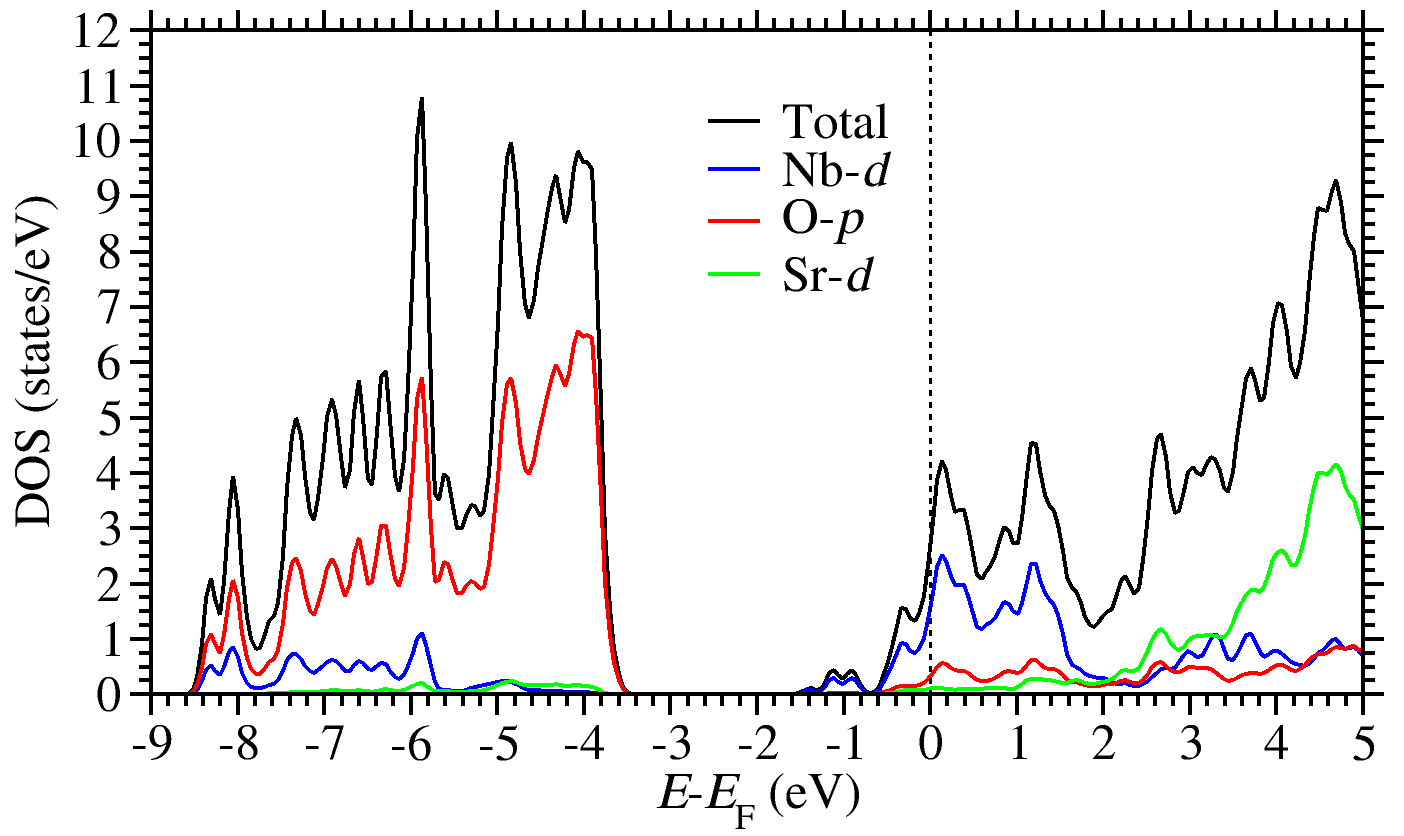}
    \caption{Sr$_2$NbO$_4$ density of states from DFT calculation. Black lines represent the total density of states, colored lines represent the density of states projected on the $4d$ (blue) orbitals of the niobium, $5s$ (green) and $2p$ (red) of the strontium and oxygen, respectively. Energy is indicated relative to E$_F$ (vertical dash line).}
    \label{Fig: DFT DOS}
\end{figure}

\begin{figure}[b]
\centering
\includegraphics[width=0.88\columnwidth]{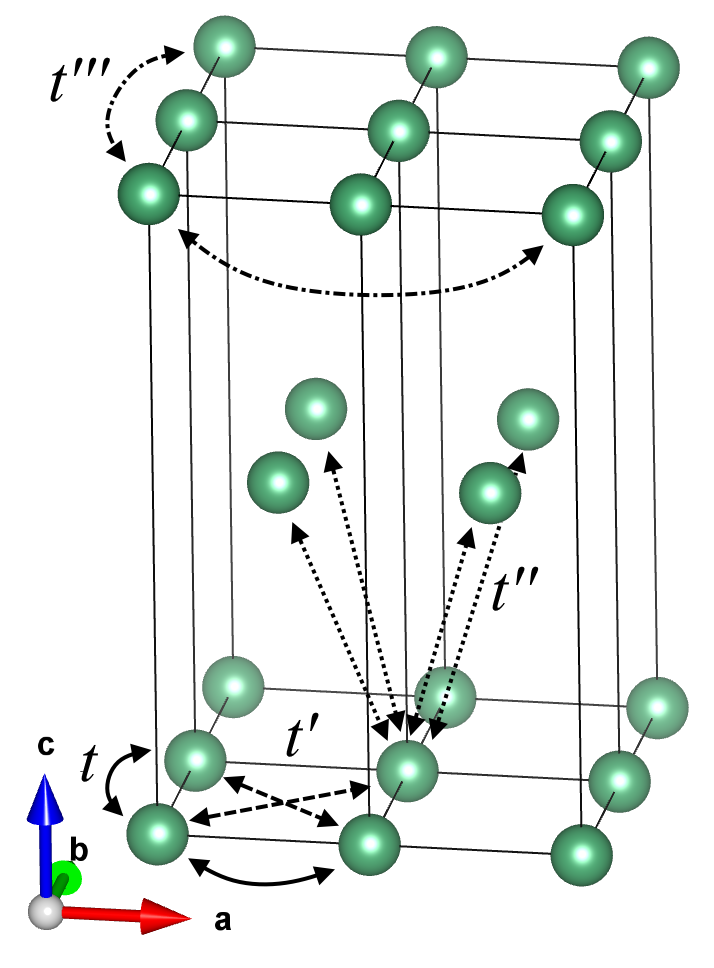}
    \caption{opping integrals in the Nb sublattice (shown for a 2×2×1 supercell of the Sr$_2$NbO$_4$): $t$ (Nearest Neighbors or NN, solid arrows), $t'$ (Next NN, dashed), $t''$ (3rd, dotted), and $t'''$ (4th NN, dash-dotted). Arrows indicate the direction of electron hopping between Nb sites.}
    \label{Fig: Hoppings Nb sublattice}
\end{figure}

\begin{table}[h]
    \centering
    \caption{Symmetry inequivalent hopping matrix elements $t^n_{i,j}$ (in meV) between Nb $t_{2g}$ orbitals ($i,j=xy,xz,yz$) for the $n$-th nearest neighbors in Sr$_2$NbO$_4$ primitive unit cell, derived from maximally-localized Wannier functions. Hopping integrals in the Nb sublattice presented in Fig. \ref{Fig: Hoppings Nb sublattice}}
    \begin{ruledtabular}
    \begin{tabular}{lcr}
                  & $d_{Nb-Nb}$, \AA & $t^n_{i,j}$, meV \\
        \midrule
        $t_{xy,xy}$ & 4.04 & 421 \\ 
        $t_{xz,xz}$ & 4.04 & 357 \\ 
        $t'_{xy,xy}$ & 5.71 & 96 \\ 
        $t'_{xz,xz}$ & 5.71 & 9 \\ 
        $t'_{xz,yz}$ & 5.71 & 7 \\ 
        $t''_{xz,xz}$ & 6.99 & 34 \\ 
        $t''_{xz,yz}$ & 6.99 & 38 \\ 
        $t''_{xy,xz}$ & 6.99 & 3 \\ 
        $t''_{xy,xy}$ & 6.99 & 1 \\ 
        $t'''_{xy,xy}$ & 8.08 & 19 \\ 
        $t'''_{xz,xz}$ & 8.08 & 35 \\ 
    \end{tabular}
    \end{ruledtabular}
    \label{table:Hoppings}
\end{table}

\clearpage

\section{DFT+U(+SOC) results}

\begin{figure}[h]
\centering
\includegraphics[width=1\columnwidth]{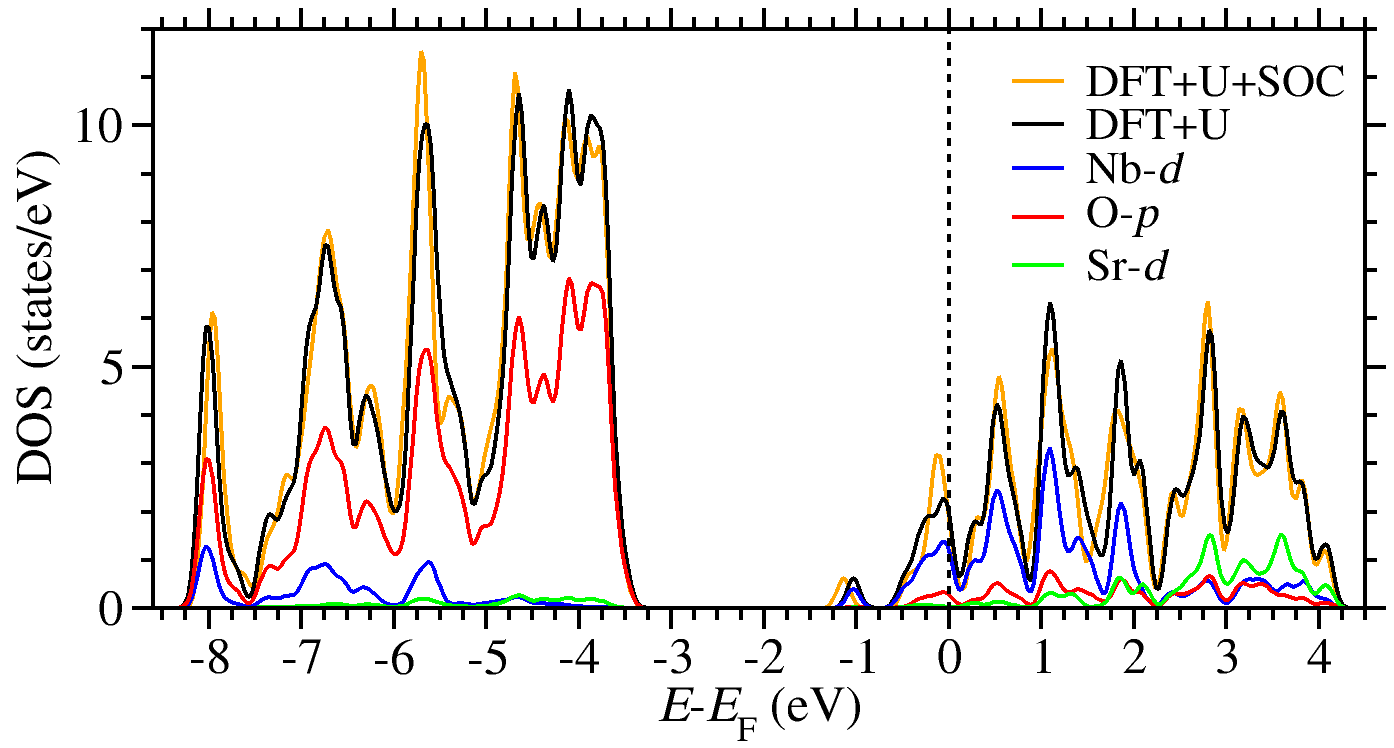}
    \caption{For comparison, the total density of states in the DFT+U (black) and DFT+U+SOC (orange) calculations of the AFM-I ground state for spin up is shown. Colored lines represent the DFT+U density of states projected on the $4d$ (blue) orbitals of the niobium, $5s$ (green) and $2p$ (red) of the strontium and oxygen, respectively. Energy is indicated relative to E$_F$ (vertical dash line).}
    \label{Fig: DFT+U(+SOC) DOS}
\end{figure}

The total and projected DOS plots for DFT (Fig. \ref{Fig: DFT DOS}), DFT+U and DFT+U+SOC (Fig. \ref{Fig: DFT+U(+SOC) DOS}) show that O-p states dominate in the region from -8.5 to -3.5 eV below the Fermi level, while Nb-d lie in the vicinity of $E_F$. Sr-d states predominantly occupy the region from 3 eV above the Fermi level.

\begin{figure}[h]
\centering
\includegraphics[width=1\columnwidth]{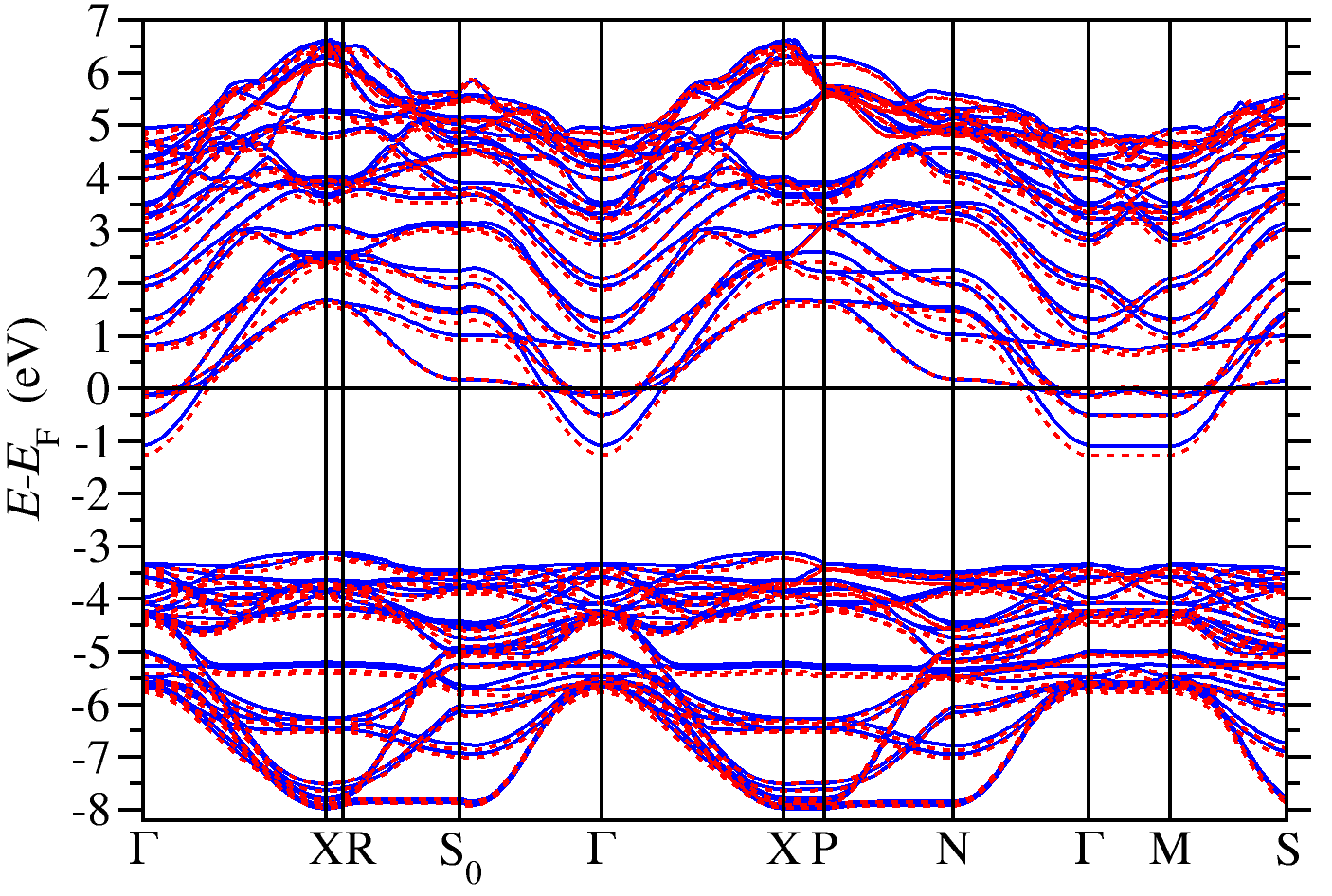}
    \caption{The electronic band structure of Sr$_2$NbO$_4$ obtained in DFT+U and DFT+U+SOC. Energy is indicated relative to E$_F$, which was set to zero.}
    \label{Fig: DFT+U(+SOC) bands}
\end{figure}